# The xDotGrid Native, Cross-Platform, High-Performance xDFS File Transfer Framework

Alireza Poshtkohi[1] , M.B. Ghaznavi-Ghoushchi [1,*]

[1] *Department of Electrical Engineering, Shahed University, Persian Gulf Highway, Tehran 3319118651, Iran*

**Abstract**

In this paper we introduce and describe the highly concurrent xDFS file transfer protocol and examine its cross-platform and cross-language implementation in native code for both Linux and Windows in 32 or 64-bit multi-core processor architectures. The implemented xDFS protocol based on xDotGrid.NET framework is fully compared with the Globus GridFTP protocol. We finally propose the xDFS protocol as a new paradigm of distributed systems for Internet services, and data-intensive Grid and Cloud applications. Also, we incrementally consider different developmental methods of the optimum file transfer systems, and their advantages and disadvantages. The vision of this paper tries as possible to minimize the overhead concerned with the file transfer protocol itself and to examine optimal software design patterns of that protocol. In all disk-to-disk tests for transferring a 2GB file with or without parallelism, the xDFS throughput at minimum 30% and at most 53% was superior to the GridFTP.

## 1. Introduction

As a fundamental result of information age, communication mechanisms and data transmission protocols provide an infrastructural foundation for the emergence and evolution of enormous computing paradigms, and integrating data access on a world-wide scale. The two open standards protocols of HTTP [1] and FTP [2] have provided basic file transfer functionalities. To overcome the problems concerned with these two protocols that are mainly due to the overheads of the TCP protocol in its window-based congestion control mechanisms used, the GridFTP protocol [3][4][5] has been proposed. In [6][7][8][9][10][11][12], we introduced a hybrid concurrent file transfer protocol, called as DotDFS, integrated with a set of event-driven and threaded-based models. DotDFS was the first file transfer protocol that, in addition to propose a new computing paradigm in the field of data transmission protocols, unveiled many architectural problems regarding the FTP and GridFTP protocols.

The TCP protocol has been used as a transport-level communication protocol on the Internet over the years. However, TCP is a rather old communication protocol designed in the 1970s. Many problems regarding the TCP have been reported such as its debility to support the increasing speeds of modern networks. One commonly-used way to reduce the overheads posed by TCP is to simultaneously choose an optimum number of TCP connections and the TCP socket buffer size, which are discussed fully in [6][7][9]. This paper tries as possible to minimize the overheads concerned with the file transfer protocol itself and to examine optimal software design patterns of that protocol. This goal plays a key role to reduce the problems associated with TCP overheads which decrease the throughput of the file transfer system. It increases dramatically the entire system efficiency and reliefs exposed drawbacks. However, it is necessary noting that the XDSI structure (refer to the Sections 3.1 and 3.2) allows xDFS protocol to operate over more optimum transport protocols (e.g., SCTP [21]) than TCP. XDSI and the xDFS framework will bring various opportunities together for research communities to implement non-TCP XDSI-enabled drivers for the sake of achieving a virtually zero-percent-overhead file transfer system currently in userspace.

xDotGrid project [6] is a new effort to establish an OS-based framework for rapid development of high-performance distributed paradigms heavily relied on the .NET ECMA standards [13][14][15]. The basic infrastructure is inspired from the mostly concepts available in Cluster, Grid and Cloud environments. From the viewpoint of the developer users, the xDotGrid.NET framework provides a set of rich, cross-platform, object-

---

\* Corresponding author.

*E-mail addresses:* alireza.poshtkohi@gmail.com (A. Poshtkohi), ghaznavi@shahed.ac.ir (M.B. Ghaznavi-Ghoushchi).

oriented, high-performance and low-level C++ class libraries which enable quick and solid development of the next-generation networked/distributed applications and paradigms. From the viewpoint of the xDotGrid core, it supplies some critical services developed with parts in kernel mode and some other parts in user space. What will be presented in this paper is not just the port of an existing code to the native code. xDFS has been designed from scratch, and this paper explains the authors' experiences to implement optimal software systems for distributed systems in a native, cross-platform and cross-language manner. In large parts of this paper, we introduce different architectures of optimal server design patterns particularly for the orientation of file transfer systems, where due to our knowledge virtually no attention has been paid to them so far. The orientation of xDFS framework is independent of technology due to its highly cross-platform manner and its architectural standards-based patterns and can be used to connect with legacy systems as well. Also, we will discuss the position of our framework among exiting parallel and cloud storage systems. Interested readers can refer to our textbook in [6] to get more familiar with xDotGrid project.

The rest of the paper is organized as follows. Section 2 concentrates on designing high-performance server architectures for Grid-based file transport protocols particularly for data-intensive Grid applications and Internet services. Section 3 describes DotDFS and xDFS file transfer protocols, and discusses new xDFS extensions over DotDFS protocol. Section 4 discusses the native and cross-platform implementation of xDFS protocol atop xDotGrid.NET Framework; moreover, making use of the concept of communicating finite state machines will help the reader more realize the presented material throughout this section. The experimental studies are described in Section 5. Section 6 determines the position of the xDFS framework among parallel storage systems. Section 7 concludes the paper and sketches our future research works.

## 2. High-Performance Server Design Architectures for Grid-based File Transport Protocols

In this section, with regard to the term high-performance server design in mind, a solid effort is employed to explain ways which specify a roadmap to performantly program server applications by developers in Grid and Cloud environments. It also makes clearer to the reader the soul and original idea behind the design and implementation of DotDFS and xDFS protocols. It can be firmly stated that the xDotGrid project in conjunction with the xDFS project is a Grid project in Grid communities where they have taken into account those technical points. Whereas web servers play a vital role in delivering Web content to users for the enterprise's business in the Internet industry, they are the most critical network servers that must deal with a large number of user requests at the same time (in typical cases as 10000 simultaneous requests or even more) and process them. Over the past two decades, extensive research in the design and implementation of optimized, stable and reliable web servers has been made. This section also outlines the taken lessons from this interesting research arena [16][17][18]. In general, four main factors affect server application performance. Furthermore, these factors impress the classification of different server designs (in this section we will refer to them). These four realistic factors are data copies, memory allocation, context switches and synchronization issues. At first glance for a better understanding of these four factors, the concepts of user space and kernel space are briefly discussed in an operating system. Most of the modern operating systems divide the accessible virtual memory to the user into kernel space and user space. In fact, kernel space is the most significant core of an operating system that is reserved to achieve efficient underlying hardware functionalities. In contrast, the user space is the area of physical memory in where all user applications are running in user mode, and this memory is returned to the system if needed. One of the kernel's roles is to manage individual user processes and prevent the interference of processes with one another. Kernel space can only be accessed by kernel space processes through either system calls or low-level APIs which encapsulate the kernel features (such as device drivers). However, the separation between kernel space and user space is to avoid data interference of these two environments and reduce operating system instability, but in this context the efficiency as a key parameter is sacrificed. Modern operating system kernels are categorized in four classifications: monolithic kernels, microkernels, hybrid kernels and exokernels [19]. In the remainder of this section, we elaborate on the four factors expressed in the design of server programs.

### 2.1. Data Copies

Eliminating unnecessary data copies can increase considerably performance of the most server applications. In the simplest case to prevent data copies, some primary methods like indirection and pass buffer descriptors (or chains of buffer descriptors) may be exploited rather than simply using buffer pointers. Avoiding data copies are sometimes very difficult in the development cycle of a server program within its source codes. For instance, in some cases which data is mapped into the user mode address space, different socket library implementations do perform one or more copies before delivering buffers to the network adaptor. Even in places that data copies are removed, additional overhead to read data and calculate a checksum remains. In fact, the main problem originated from the data copy is due to extra copies from user space to kernel space and vice versa. Traditionally, the kernel has provided a layer of abstraction between applications and hardware, and also has been responsible to exchange data between them. This way requires two additional data transfers from an application program to the kernel and from the kernel to the hardware, compared with the scenario that the application program could have directly access to the hardware if needed. Moreover, this relation between hardware and software allows to use DMA operations that will relieve the CPU, but such a capability does not exist between two pieces of software (i.e., between an application program and the kernel). A method called zero-copy enables such a feature. Zero-copy techniques fall into three categories as follows:

*1. Data transfer optimization between kernel and application programs:* this method is on the optimization basis of CPU copies between kernel and user space in which the traditional methods in classifying the communications are maintained and a more flexible approach is achieved.

*2. Avoidance and optimization of in-kernel data copies:* this class of techniques is going to implement new system calls or optimize traditional methods to achieve more performance in certain cases that data can be fully processed in the kernel.

*3. A byway on the main data processing path:* with regard to the method 2, the kernel sometimes has no need to meet directly with data, and it can be avoided. On the other hand, in this category of techniques, they allow the direct data transfer between user space memory and hardware, and the kernel just manages transfer operation.

## 2.2. Memory Allocation

Memory allocation and de-allocation are two of the most important operations among long-running server programs. Two types of memory allocators exist called as custom and general-purpose allocators. Many of the general-purpose memory allocators have been implemented for C and C++ languages. These allocators create a good running time and low fragmentation for a wide range of applications. However, the use of customized memory allocators can take advantage of application-specific behavior. They can dramatically increase performance. Custom memory allocators can benefit from specific allocation patterns with many operations at the lowest level cost. For example, a programmer can make use of a region allocator to assign a number of small objects with a known lifecycle and frees all of them at a given time. This typical custom allocator returns individual objects from a range of memory, and then the whole region is de-allocated. To attain high performance, programmers often develop their own ad hoc custom allocators as macros or monolithic functions (like inline functions) so as to avoid function-call overhead. In fact, these methods to improve performance have been recognized as the best habits of skilled computer programmers. Generally, the requirements of a dynamic memory allocator system can be summarized as follows below. *Stability*, it is necessary to contiguously keep stable the memory allocator system performance for long-running server programs. Obviously, the throughput of such a system must remain stable over time. *Speed*, such a system must be fast as possible in memory allocation and de-allocation. A memory block should not depend only on the thread that has allocated it. Threads must be able to operate collectively on a shared allocated memory area. This factor makes possible to pass the object references among different threads. From other important points in the design of memory allocators can be noted to scalability, size independency and maximum locality.

## 2.3. Context Switches

Context switching is used as the basic mechanism to share a processor among multiple threads of execution. Each thread is dependent on general-purpose registers, status registers and a processor state such as the program counter. A context switch is an operation to save the process state of a thread and load another thread, particularly in hardware-implementation context switching at the pipeline stages of a follow-on chip multi-threaded (CMT) processor (e.g., Fetch, Thread-Switch, Decode, Execute, Memory, and Writeback). If threads relate to different virtual address spaces, a context switch also contains switching the address translation mappings used by the processor. Switching the address space requires that the relevant inputs in the process's address translation cache (TLB) are invalidated. If the instruction or data caches are tagged using virtual memory addresses, they will have to be emptied as well. Context switching imposes a small performance penalty on threads in a multithreaded environment. In addition to direct overheads concerned with a real context switching code, there are numerous other factors that contribute to the overhead penalties. Another indirect overhead is due to disorder in branch-target buffers, and CPU cache like instruction set, data and address translation. However, another source of these indirect overheads may be operating system memory paging. A context switch can result in an in-use memory page being moved to disk if there is no free memory, thus hurting the total performance. Context switches can take place in the kernel. Kernel mode is a privileged mode of CPU, in which only kernel code is executed, and provides access to all of memory locations and other operating system resources. Other programs, including applications that primarily are executed in user mode can run parts of the kernel by system calls. The existence of such a structure is considered as a mode switch (or mode transition) instead of a context switch, because this structure does not change the state of the current process or running thread. So a context switch is used as a mechanism to switch between two threads of execution. Therefore, we infer that a system call is not actually a context switch; indeed, it is akin to a simple function call that causes to change state of the processes from unprivileged user mode to a privileged kernel mode. Memory mappings are not switched. Also, the return of mode transition to userspace from kernel during returning from a system call is similar to the return operation of a userspace function call. In addition to context switches occurred between threads in software level, in hardware a processor interrupt causes the state of running task to be saved, while an interrupt service routine is executing. When interrupt service routine is completed, the saved state will be restored again. While memory mappings are not actually switched during interrupt servicing, it does becloud the cache state and may also constitute some indirect overhead. Hence, context switching shows a substantial cost to the system in terms of CPU time for a typical operating system.

## 2.4. Synchronization Issues

Programming complexity is an ambiguous issue in writing multithreaded applications with shared memory regions. Although threads simplify the design logic of programs as possible, great skill and experience are required to ensure that the correct relationship among threads has been established. Errors for the selection of appropriate synchronization methods among a set of threads while accessing to shared objects result in an incorrect execution of programs. These methods are very sensitive in most cases. In many programming languages, locks are essential synchronization constructs to enforce limitations for having access to a resource in a concurrent environment in which many threads of execution exist. Two key constraints must always be considered in using locks:

*1. Performance:* A complex trade-off often exists between programmability and performance because most programmers have to make their decisions on how to share data during the code development process using static information for dynamic runtime behavior. Programmers usually use conservative synchronization to write correct codes and keep them simple. While such this use can guarantee correctness, create stable software and lead to faster code development, but it prohibits parallelism. Fine-grained locking may help improve performance, but they make the code hard and error-prone to write. Coarse-grained locking may facilitate to write suitable and good code and reduce errors, but hurts to the key factor of efficiency. In addition to these problems, locks can impose very important overheads, serialize the execution of programs and reduce the overall system performance.

*2. Stability:* If a thread acquires a lock and marks it as held, other threads acquiring this lock must wait until the lock is free. Such wait can implicitly influence on the system behavior being designed. If the lock owner is de-scheduled by operating system, other threads waiting for this lock cannot continue their execution since the lock is not free. If the lock owner aborts, other threads waiting for this lock never complete; hence, this lock is never free. As seen in this scenario, the shared memory regions by abnormal termination of a thread remain in an inconsistent state. This causes critical sections to be held in a messy granularity.

Widely-used, general-purpose locking mechanisms include mutexes, semaphores, condition variables and multiple readers and single writer locks. Other major problems caused by locks are lock contention (due to excessively coarse granularity or inappropriate lock type), deadlock (each thread of control is waiting for a lock held by another thread of execution), lost locks, race conditions, and incomplete or buggy lock implementation. In total, the general overheard associated with locks can be summarized in extra resources for using locks akin to the memory space allocated for locks, the time needed for acquiring or releasing locks, and the CPU time to initialize and destroy locks. Therefore the more locks a program uses, the more overhead associated with the usage.

## 2.5. The Proposed xDFS Server Architectures in FTSM Upload Mode

With lessons taken from four factors explained in the design of a high-performance and stable server, we can divide the existing, widely-used server architectures into six main kinds including multiple-process architecture, multiple-thread architecture, single-process event-driven architecture, multiple-process event-driven architecture, multiple-thread event-driven architecture, and staged event-driven architecture (SEDA) [20]. Each of these architectures has advantages and disadvantages in designing every given application-specific server. One of the chief novel contributions of this paper is to extend these architectures relied upon the inherent structures of the DotDFS and xDFS file transfer protocols. In all these models, we consider a file transfer scenario as xDFS or DotDFS FTSM Upload mode that is in progress from one client to the server using *n* parallel connections per transfer session. Although these models have their own novelty suggested by the authors of this paper, it is worthwhile to note that, as stated in Section 4, the core of xDotGrid server has conceptually been made upon hybrid server architecture. Despite these models offer an abstraction for server-side protocol implementations, it is also required to note that all implementations of client-side APIs have benefited practically from these quasi-server-side architectures for designing a whole real-time, high-performance client-server system. Because of lack of space, we omit to describe the details of client-side implementation in this section. As follows, we suggest and explore three major models more suitable to design high-performance servers in the areas of file transfer protocols and file servers. The first model has classically been used extensively (such as GridFTP server), but the second and third architectures are proposed completely by the authors of this paper.

### 2.5.1. Multi-Processed xDFS Server Architecture

Let's start with a familiar case: a client intends to upload a large size file to the xDFS server through the well-known *n* parallel TCP streams using FTSM mode. As shown in Fig. 1, in the multi-processed (MP) model, a process called *acceptor process* gets the new connections inside the body of the *main()* function. Each client request gets mapped into a process which manages the TCP stream. Fig. 1 illustrates the set of *n* processes to representing one xDFS FTSM session. Process 1 to *n* may be retrieved from a process pool, or if there does not exist either enough or idle processes in the process pool, then the *acceptor process* can call the system functions, *POSIX fork()*/Win32 *CreateProcess()* , which create a new process to manage the new connection. Synchronization can be challenging in this MP model, because each of the processes are executed in a separate address space. This problem can be resolved by using an IPC (Inter-Process Communication) technique. This IPC mechanism, for example, may be used for passing the client socket handles to the processes, and for synchronization among multiple processes, the *main()* function and the process pool. As it can be derived from Fig. 1, the MP model imposes three major overheads to the system including large opened file handles, heavyweight context switches, and excessive off-chip/on-chip memory used.

*1. Large opened file handles:* In the MP model, a file handle is opened per process. The set $\{fd_1, fd_2, ..., fd_n\}$ represents the opened file handles for *n* parallel connections. Clearly, these *n* processes are concurrently receiving file block from the client. In fact, a single file through *n* separate file handles is shared to be written by the system function *write()* among multiple processes. From two perspectives, this model can decrease significantly the file transfer throughput of a single xDFS FTSM session. One negative factor on performance is the nondeterministic distribution of random disk *seek()* operations among *n* parallel processes, it is also necessary to note that each operating system according to its underlying I/O scheduling algorithms and policies usually behaves typically differently on the disk I/O throughput. Disk schedulers in current operating systems are generally work-conserving, i.e., they schedule a request as soon as the previous request has finished. The overhead of acquiring and releasing each lock in the OS kernel may be considerable on the overall disk I/O throughput. File system performance is often a major component of the total system performance, and in this case is heavily dependent of the nature of server application operating the workload. In fact the use of MP model with the large open file handles can cause four major performance penalties. It increases the number of I/O's to the underlying device(s). It violates the grouping of smaller I/O's together into larger I/O's where possible. It cannot be used to optimize *seek()* pattern to reduce the amount of time spent waiting for disk *seek()* operations, the disk *seek()* operations are expensive head repositioning operations. Finally, it is not possible to cache as much as data as realistic to reduce physical I/O's. Perhaps at first glance, it seems clear that these problems can be avoided in the following method: an extra process called the *disk I/O process* have access to a single file handle, to individually coherently write file blocks to the disk, and all other processes have access to the *disk I/O process* by an IPC mechanism. However, this technique implicitly explains that the use of an extra process, an IPC mechanism and synchronization at the process level in user space, eventually degrades system performance.

*2. Heavyweight context switches and excessive used off-chip/on-chip memory:* As sketched out in Fig. 1, two of the coarse-grained approaches to creating new system processes are used including POSIX *fork()* and Win32 *CreateProcess()*. Though, it may apparently appear that an infinite number of processes can be created. Even so, it should be mentioned that operating systems only permit a limited number of processes to be assigned within their available amount of physical memory. Even though, using a process pool cannot be appropriate for an xDFS server in crucial applications including fine-grained parallel or highly data-intensive programs in Grid environments, and in high-traffic environments with too many clients' connections like the Internet. Approaches to making a process structurally are divided into *creation* and *clone* modes. In the *creation* mode, the operating system performs the following operations: loads code and data into memory, creates an empty stack, initializes state to same as after a process switch, and makes process ready to run by inserting into OS scheduler queue. In the *clone* mode, the operating system performs the following operations: stops current process and stores its state (it means transiently freezing the entire application especially in the case of a server software program); makes copy of current code, stack and OS state; and makes the new born process to be run. Forks are in the *clone* mode while processes made by calling the system function *CreateProcess()* are in the *creation* mode. As it can be seen in the MP model, processes impose two other important overheads including the excessive used off-chip/on-chip memory, and heavyweight context switches due to the use of *n* processes. Also, process context switch implies getting a new address space in place by page table and other memory mechanisms.

### 2.5.2. Multi-Threaded xDFS Server Architecture

Choosing the appropriate thread model for server programs is a complex decision influenced by many factors, including performance constraints, software maintainability, and the presence of existing code. To some extent, for reducing the process-based overhead of the MP model, multi-threaded (MT) architecture is proposed for xDFS server in FTSM mode. The MT structure is shown in Fig. 2. In the MT model, processes are replaced with threads. These multiple kernel threads share a single address space and are accessible to all threads. Each thread manages one stream from the remaining *n-1* streams. In this model, threads can share all public information, it makes possible to remove IPC mechanism used in the MP model and allows the *thread acceptor* directly to manage the running threads. Threads are lightweight processes and expose much less overhead than the MP model on server applications. This model substantially reduces the total memory (physical and virtual or swap space memory) used by programs. To eliminate the overheads due to large opened file handles, in the proposed MT model, a thread named the *disk thread* is used to manage *write()* and *seek()* operations for received file blocks from the client. In Fig. 2, the *disk thread* only opens a single file handle from the requested file. File blocks are put into a circular buffer which contains the most recently file blocks received from the client. To avoid race conditions a pessimistic locking mechanism is used. The thread that tends to put a received file block into the circular buffer first acquires the buffer lock, and then independently takes action to fill the buffer. The *disk thread* attempts to arrive at a relative coherency and reduce the number of disk *seek()* operations through a scatter/gather I/O, a.k.a vectored I/O, mechanism to contiguously write the content stored in the circular buffer into the storage system at a whole. This buffering method can significantly decrease many successive calling the function system *seek()* for performance enhancements. Further, the *disk thread* has the duty to inform the xDFS client of correctly and errorless receiving of the file block via sending an exception header defined in xDFS protocol. The MT model can have too much few overhead in comparison to the MP model, but this model still imposes two critical performance pitfalls. In this model, due to the *n+1* threads' existence for each xDFS session once more leads to many context switches, although this overhead is much less than the MP model.

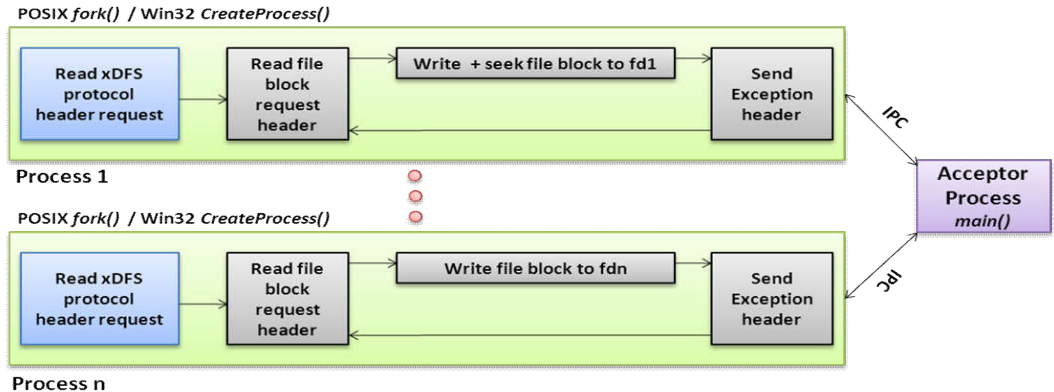

Fig. 1. Multi-processed xDFS server architecture.

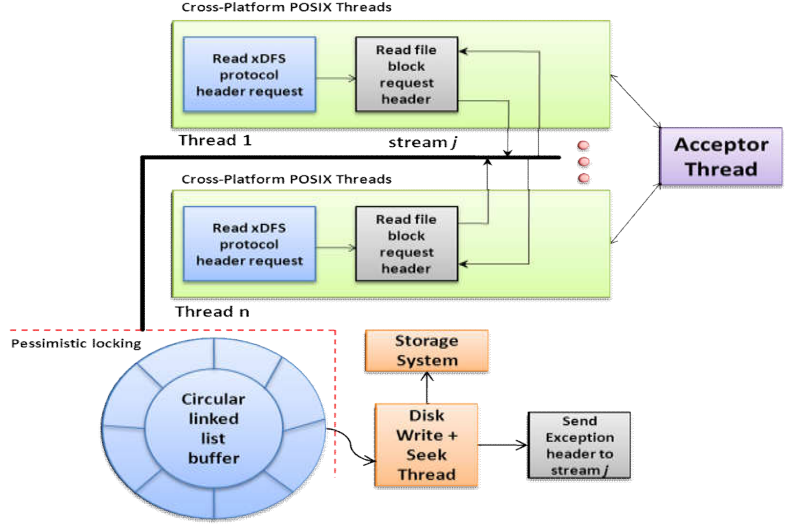

Fig. 2. Multi-threaded xDFS server architecture.

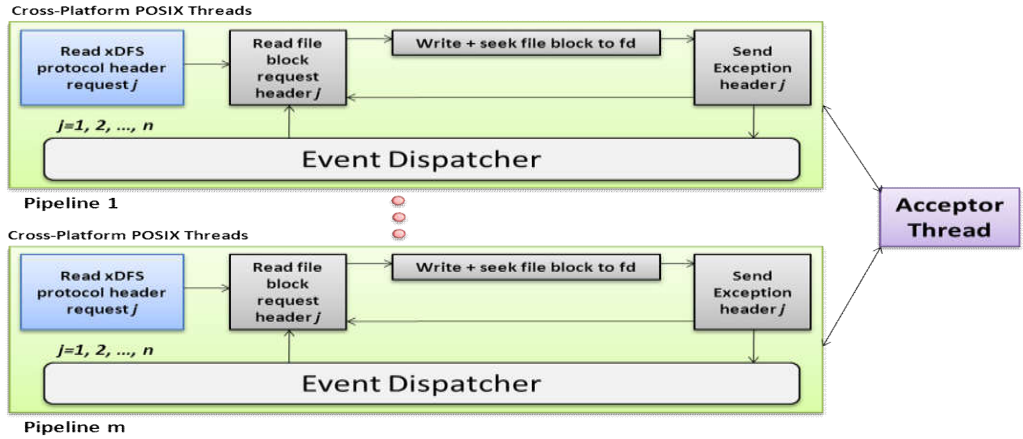

Fig. 3. Multi-threaded event-driven pipelined xDFS server architecture.

TABLE 1: THE NUMBER OF THREADS

| (1) | (2) |
|---|---|
| $T_{MT} = \sum_{i=1}^{m}(n_i + 1) = \frac{m \cdot (m+1)}{2} + \sum_{i=1}^{m} n_i$ | $T_{MTEDP} = m$ |

m is the number of server-side FTSM upload sessions.

The second overhead is the use of a pessimistic locking mechanism to synchronize threads for simultaneous access to the circular buffer, because if *n* threads are concurrently ready to write to the buffer, then one thread can only acquire the buffer in an OS time quantum. Therewith this procedure explicitly depicts that many context switches are unintentionally disparately happening among the disk thread, and the xDFS session threads which had received file blocks, but they have not made dirty the buffer yet. If an inappropriate, non-optimal pessimistic locking algorithm is used, it may frequently postpone the execution of threads, and finally reduces efficiency drastically. For instance, during the development of primary DotDFS client-side prototypes, the system performance decreased up to 50% with an improper locking algorithm used. We spent a lot of time to develop and optimize an alternative algorithm. It is also important to note that all operating systems do not support kernel-level threads. This means that cross-platform applications have to consider employing userspace-level multithreading libraries in their C macros or C preprocessor directives, like GNU Portable Threads library. This technique would lead to high synchronization overheads for sharing data among a large number of threads. There are multiple threading variants for different application-specific scenarios. In *1:1* (kernel threads) method, threads are created by a *1-1* mapping onto a single kernel-level scheduled thread. This model is supported almost in all the operating systems. In *N:1* (userspace threads) model, to the number of *N* user-land threads are mapped into a schedulable kernel thread. This method benefits from fast and low-cost creating threads. But the most fundamental pitfall of this model is that if one of the userspace threads blocks the kernel thread due to a blocking situation (such as in disk I/O-bound cases), then all other threads being able to run will be blocked. In *N:M* (scheduler activations), to the number of *N* userspace threads are mapped to the number of *M* kernel threads. As a whole the *N:M* model has too few spread among operating systems in contrast to other models, because an *N:M* library implementation requires extensive changes to both kernel and user space code.

### 2.5.3. Multi-Threaded Event-Driven Pipelined xDFS Server Architecture

In this section, we propose a multi-threaded event-driven pipelined (MTEDP) architecture for xDFS server operating the FTSM mode in which multiple pipelined apartments are overlapped in execution. To increasingly enhance the performance of xDFS server, MTEDP completely eliminates the synchronization mechanisms used in the MP and MT models and reduces the number of context switches to a large extent. Fig. 3 shows the MTEDP architecture. Due to similarities in Fig. 3 with the pipelining techniques in computer organization we included the term *pipelined* in the phrase of MTEDP. These *m* pipelines, in which each pipeline contains *n* parallel connections, actually indicate *m* parallel file transfer sessions in FTSM mode. Each pipeline manages one transfer session. In this model each pipeline in every thread owns *n* socket handles and these handles, to asynchronously send and receive data over sockets, are managed through event dispatching and multiplexing techniques realized as a collection of comprehensive communicating fine state machines (CFSMs). As it is obvious each pipeline has an opened file handle and this would lead to avoid the use of pessimistic locking mechanisms, which reduced the performance in the MT model, and the problems of large opened file handles in the MP model. Now, we can derive the relations in Table 1 to representing the number of threads created in an xDFS server for both the MT and MTEDP models, in where *m* is the number of FTSM transfer sessions, and each of which has $n_i$ parallel connections. There are different event dispatching and multiplexing network I/O mechanisms. However, some of them are implementation-specific to some operating system platforms and each of them has its particular advantages and disadvantages. In general, these mechanisms can be split in four major categories.

*1. select() and poll():* The system-calls *select()* and *poll()* are stated-based event dispatching mechanisms. They report the current status of a set of sockets as their input arguments. When there are a large number of sockets, *select()* is more suitable because less data is copied to or from the kernel. The *select()* function is available on the most platforms, but *poll()* has fewer spread (for example, *WSAPloll()* has been added to the Windows Vista operating system and its later versions). For this reason, the implementation of the both xDFS client and server cores relies upon the *select()*.

*2. POSIX.4 Real Time Signals (RT Signals):* RT signals are an extension to the traditional UNIX signals. They allow the kernel to queue multiple instances of a signal for one process. Linux kernel 2.4 extends RT signals so that the opportunity for delivery of socket readiness is provided by a particular real-time signal. RT signals are not available on all platforms. One of the main disadvantages of RT signals is that they make the server code complex to write in contrast with other methods. Also, because the system call *sigwaitinfo()* is used to implement RT signals, the switching between kernel and user modes take places many times. The main advantage of RT signals over against *select()* and *poll()* is that they are much more scalable.

*3. /dev/poll*: */dev/poll* appeared for the first time on Solaris 7 for removing the need to specify the desired set on every *poll()*. This way is a state-based event dispatching mechanism. The main idea behind this method is that application programs can open device files into the kernel to make a set of favorite descriptors. This set is gradually made after accepting a new connection from the network adaptor. This method can reduce the amount of favorite copied sets between user space and kernel space. The main drawback of this method is that it is limited to UNIX-like operating systems and is not supported by Windows family of operating systems.

*4. NT I/O Completion Ports and POSIX AIO:* NT I/O completion ports supply an efficient threading model for processing multiple asynchronous I/O requests on a multiprocessor system. When a process creates an I/O completion port, the system creates an associated queue object for requests whose sole purpose is to service these requests. Processes that handle many concurrent asynchronous I/O requests can do so more quickly and efficiently by using I/O completion ports in conjunction with a pre-allocated thread pool than by creating threads at the time they receive an I/O request. Completion ports are only available in Windows platforms. The POSIX AIO interfaces allow a process or thread to start multiple simultaneous read and/or write operations to multiple file descriptors, to wait for or obtain the completion notification of requested operations, and to retrieve the status of completed operations. One of the current AIO's disabling downsides is that they are not applicable to the Linux kernel-mode AIO for network I/O subsystems (e.g., sockets) and also are not available in Windows platforms.

## 3. DotDFS and xDFS File Transport Protocols

DotDFS and xDFS are general-purpose network protocols to achieve the goals of high-throughput file transfers and network file systems. DotDFS protocol was proposed based on the demands of Grid communities. xDFS investigates new objectives beyond its DotDFS predecessor. xDFS adds so many new extensions to the DotDFS protocol to facilitate much more use of xDFS protocol in other dynamic environments such as Cloud Computing and the Internet. The xDFS protocol allows multiple clients to have simultaneous access to managed files and directories (for large or small size of volumes with high-throughput performance) hosted on desktop, dedicated server systems or any other computing entity. Additionally, it makes possible to access other services including interprocess communication, remote file streaming, and authenticated transports over all xDFS channels. Totally, xDFS is a client-driven protocol in which a client makes a request to the server and server replies to the client. xDFS is both a stateless and stateful protocol. It imposes states to maintain security contexts and cryptographic mechanisms via xSec (DotSec) protocol, and file access semantics such as locking. Furthermore, xDFS protocol defines a set of specifications to achieve the aggregating throughput of widely-used TCP protocol in WAN and MAN networks for file transfers. It enables multiple clients to simultaneously share files on server systems. It ultimately leads to facilitate collaboration, and efficiently use and centrally manage resources. DotDFS protocol introduced three operating modes including FTSM, DFSM, and PathM. DotDFS, with three these modes, attempted to accommodate itself to the expressed needs above [6][7][8][9][10]. The xDFS protocol in a new look to these three modes tries to highly extend DotDFS features in more accommodation with Internet services. In this paper FTSM mode is fully considered. However, we exclude to describe xDFS extensions on DFSM and PathM modes, because they are our future research work and are outside the scope of this paper. Section 3.1 discusses overall xDFS features, and Sections 3.2 and 4 take a comprehensive description of xFTSM protocol.

### 3.1. Overall xDFS Features

*Transport independence:* xDFS protocol does not necessarily require the use of any specific network transport protocol. TCP is the default scheme of a connection-oriented transport protocol used to carry xDFS binary headers. As a whole, this flexibility is due to the layered model of xDFS specification. xDFS uses xDotGrid Socket Interface (XDSI) architecture for underlying transport protocol in the lowest layer. For example, the highly extensible XDSI architecture allows developers to implement xDFS over a variety of transport stacks such as SCTP [21], UDT [22] and RDMA [23] with the minimum changes in xDFS C++ source codes. Since the main audience of xDFS protocol is various purposes in the Internet, the major part of this paper discusses an implementation of xDFS over TCP-enabled XDSI.

*Flexible connectivity:* In xDFS protocol, a single client can connect to multiple servers, or it can establish one or more connections on each server. The activity of multiple client processes can be multiplexed over a single connection. In fact, this feature represents the support of reusable channel mechanisms in xDFS protocol.

*Feature negotiation and prerequisites:* Because the collection of xDFS protocols always during the coming years will be evolving, the feature negotiation was added to xDFS. This feature provides the negotiation of dialect and the supported feature set of the protocol between two endpoints. For instance, before connection establishment between client and server the protocol version is talked used for more interoperability and the support of legacy applications. As another example, the negotiation of per-connection basis is used to choose the type of transmission channel.

*Resource access:* A client can simultaneously access to multiple resources shared across remote computing entities. Additionally, a client can have access to files and directories for different purposes. As a unique feature, the support of named pipe inter-process communication (IPC) was added to the set of xDFS protocols. A client can open, read, write and close named pipes on the target server. Named pipes can be used as a communication path between client and server processes. This feature enables the possibility of using xDFS protocol in parallel computing applications based upon a concept of distributed remote file access.

*Unicode file name support:* Since xDFS protocol extends DotDFS protocol and is implemented based on native C++, it uses the features of xDotGrid.NET Framework classes for string manipulation. The class *System::String* supports the default format of Unicode strings. All the xDotGrid Framework's methods and functions make use of this primitive data type. They are redirected to the native Unicode strings before having access to call Win32 or POSIX APIs. This feature not only has no overhead, but also it leads to more xDFS multilingual universality.

*Distributed File System Mode (DFSM):* This requested mode supports the access to files and data sharing mechanisms, which have employed in conventional distributed file systems. Additionally, this mode can be used for stripped and third-party data transfers. One good example of this mode is a situation with one or more transport streams between *m* network endpoints on the sending side and *n* network endpoints on the receiving side.

*Path Mode (PathM):* The design goal of PathM mode is to support basic features like creation/deletion of remote files/directories and related features. In PathM mode, the xDFS server operates like a RPC server, but all the methods requested by the client are previously defined as binary in the client-server negotiation protocol.

*Authentication, data integrity and data confidentiality:* DotGrid and xDotGrid projects [6][10][11][12] are architectural constructs implemented as layered frameworks and are executed on the top of XDSI. xSec (DotSec) is the layer above XDSI and provides secure communications between endpoints based on xSec (DotSec) TSI. xSec is a lightweight grid security infrastructure (GSI) [7]. It is designed as a unified security model in xDotGrid platform. xDFS protocol requires all channels MUST be authenticated according to services provided by xSec security layer. xSec the new version of DotSec protocol is being documented to be implemented in native code. After authentication/authorization steps, all channels are encrypted if the higher layers or application programs request it from the xSec layer.

**3.2. xDFS xFTSM Protocol**

The xDFS protocol is more sophisticated than the FTP and GridFTP protocols (and even DotDFS protocol) in terms of structure and extensibility. The xDFS using a fully binary protocol model and requiring the use of comprehensive finite state machines precisely defines a wide range of mechanisms and operations. This approach complicates the feasible implementations of xDFS protocol, but ultimately increases significantly the performance and throughput to those systems that have been developed atop the xDFS framework. Primarily, DotDFS with the introduction of the File Transfer System Mode (FTSM) sub-protocol tried to suggest a new file transfer paradigm to solve a set of problems. They were to support parallel connections and negotiate TCP window size between a client and server. These two methods relieved partly the problems due to making TCP protocol as the widely-used transport protocol layer in the OSI model when transferring files with large sizes in order to increase the network throughput in high-latency WAN networks. xDFS protocol extends FTSM mode and proposes xFTSM architecture. xFTSM mode inherits all the properties of FTSM, changes the former FTSM structure and adds new extensions to it. Today, the design of state-of-the-art network protocols is an art of engineering. However, a standard protocol is assigned to allow different implementations to interoperate. Therefore, a standard protocol should summarize the operation of its feasible implementations. The selection of multiple implementations and many other engineering details often make the formal specification of a protocol difficult. Lack of formal specification as seen in the process of developing the IETF standards has two important negative results: the protocol accuracy is not easily verifiable, and the protocol may be ambitious in some aspects. First, the bugs are continuously identified and are resolved in the standard protocol. Second, the protocol ambiguities can open adequate space for bugs and even attacks. These bugs and ambiguities are identified in an ad hoc way. Third, protocols may be used without the verification of accuracy. According to these points and that xDFS has become more complex than its previous versions, there MUST be made use of more powerful methods than commonly methods employed for documenting the IETF standard protocols. We widely utilize communicating finite state machines (CFSMs) to propose and design xDFS protocol. Several examples of CFSMs will be mentioned in Section 4. The formal model of a CFSM plays an important role in three various areas of network protocol design: formal validation, protocol synthesis, and conformance testing. Protocol formal validation is a powerful technique for automatically checking that a collection of communicating processes in the CFSM is free from concurrency-related errors. Protocol synthesis is used to derive an implementation level's protocol specification from service specification. Protocol conformance testing is a kind of testing where an implementation of a protocol entity is tested with respect to its specification for efficiency or interoperability purposes. To precisely understand the structure of xDFS protocol, we present a preliminary introduction of communicating finite state machines (CFSMs), because we will take advantage of CFSMs to describe xDFS. Over the past thirty years, a variety of formal models have been proposed and studied to facilitate the specification and validation of concurrent systems. One major example is communication protocols in where protocol entities interact with each other in accordance with a set of rigorous rules. Designing concurrent systems is known as a significantly deep problem. One of the major sources of difficulties with concurrent systems is due to the fact that the function of these systems tends to become very complex and too large. To describe a concurrent system is difficult to the way that it operates in an environment over a period of infinite time. Functional behavior of a system is defined by the large number of the ongoing system interactions with its environment, and these interactions often show complex interdependencies from themselves. Therefore, it is very difficult to describe, understand and predict the behavior of these concurrent systems, and finally to examine whether their needs are met or not. A suitable model for describing communication protocols and concurrent systems are CFSMs. In the CFSM model, a protocol is defined as a collection of processes (i.e., the protocol entities) which exchange messages over error-free simplex channels. Each process is modeled as a finite state machine (FSM) and each simplex channel is a FIFO queue. A protocol state is composed of a state for each FSM and content for each simplex channel. A state transition occurs only when the process is ready to send a message to one of its output channels or receive a message from one of its input channels. The CFSM model is an elegant and well-defined structure, and somewhat easy to understand. These features have made it very attractive for industry and academia. Obviously, the CFSM model virtually has become the de-facto standard to specify, verify and test communication protocols in the telecommunication industry.

After this we assume that a client intends to connect to an xDFS server through $n$ parallel channels. A channel in xDFS protocol is an abstract concept during which the interaction of operation flow and data flow takes place between network endpoints. xFTSM and xPathM modes flow through these stateful channels. Furthermore, because the architectures of xDFSM and xPathM modes are beyond the scope of this paper we don't discuss them in this section. These $n$ parallel channels for xFTSM mode over TCP pipes are the same mechanisms of parallel TCP connections used to increase the throughput in wide-area networks. These $n$ parallel channels for xPathM mode can denote $n$ routes on the side that have been requested from it, they can improve the system efficiency for transferring directory trees specially in the case of a lot of small files. The implementation of xPathM mode may be taken advantage of multiple threads to handle parallel channels, which can contribute dramatically to the performance in terms of optimum use of threading concurrency. However, it is recommended strictly to limit the number of created threads equal to the number of available processor cores in xPathM mode so that the additional overhead as possible can be avoided due to context switching. To achieve the maximum performance of the proposed protocol in xFTSM mode, xDFS requires the client-server developers to use event dispatching and multiplexing methods for managing parallel network I/O in which the client or server side MUST create one thread per session. The proposed model due to compatibility with this requirement compulsorily relieves different xDFS implementations. In the rest of this paper, we assume that a client and server are negotiating to upload and download files with large sizes over xFTSM channels. It must also again be noticed that whereas channels are a general concept in xDFS protocol, they can be built over different transport protocols utilizing the underlying XDSI stack such as TCP, SCTP, UDT and RDMA. Fig. 4 shows the client-server xDFS protocol sequence diagram in xFTSM and xPathM modes. A client chooses the xFTSM mode after connecting to the server, selecting xDFS service and authentication in step 5. The steps 1 to 7 are repeated for all $n$ parallel channels involved in this process. The first client channel connected to the server is responsible to register a new xFTSM transmission session at xDFS server. This channel-based session registration MUST be performed in the stage of negotiation protocol by a unique session identifier. The most important parameter passed in this step is the number of parallel channels $n$. Table 2 illustrates some parameters in the negotiation protocol. The data structure of negotiation protocol (and any xDFS structure or object that needs to be transferred between endpoints) is transformed into a native binary format using xDotGrid Object Passing Interface (XDOPI) that can be retrieved to the initial form. In xDotGrid.NET Framework, serialization is the low-level process of converting the state of an object into a form that can be persisted or transported. XDOPI can be easily implemented in many of languages and platforms. After completing the session registration step by the first client channel, server MUST wait until other remaining $n$-$1$ channel(s) are established. After getting all the channels established, there are $n$ duplex channels between client and server which, on the demand of client, could lead to the parallel data transfer in each type of upload and download modes. One of the new and unique extensions in xDFS protocol is the abstraction of channel events added to the duplex channel negotiation as shown in Fig. 4. This way in all channels the client or server can change the operating mode during the xDFS session from xFTSM to xPathM or vice versa. To further clarify this new extension, it is necessary to describe the structure of message interchange formats exchanged in the duplex channel negotiation to the form of headers between client and server. DotDFS protocol called the term *data transfers* to the step 10 of Fig. 4. Nevertheless, xDFS protocol changes the term *data transfers*, refers it to as the term *duplex channel negotiation*, and provides the feasibility of operation flow, data flow or a combination of both through the duplex channel negotiation. This evolutionary extension on xDFS highly expands the protocol in terms of functionality and extensibility for development of the future xDFS Framework versions. An illustration of a general xDFS protocol channel binary header encapsulated in xSec TSI header appears in Fig. 5 during the step 10 of Fig. 4. The description of xSec TSI Header is beyond the scope of this paper, which encrypts all xDFS channels. Channel event represents the structure of channel headers and generally describes the operation flow and data flow in the step 10 of Fig. 4. Because the channel event exhibits high structure complexity in the xDFS specification, the pattern of xDFS channels MUST be characterized in finite state machines. Some types of channel events are shown in Table 3. The structure of a channel header related to xFTSMD and xFTSMU types is shown in Fig. 5. xFTSM header stores the information of data file blocks that are being transferred (such as the file block offset and block length). In xFTSMU and xFTSMD types, depending on the data flow, a set of operations are executed at client-server sides as follows: reading from local storage and sending to the remote server (in upload scenario initiated by the client), and receiving from the remote server and writing to local storage (in download scenario initiated by the client). Implementations to satisfy the support of various types of channel events MUST be considered as a collection of FSMs in the level of protocol and source codes to reduce complexities as much as possible.

TABLE 2: THE PARAMETERS OF NEGOTIATION PROTOCOL

| | |
|---|---|
| Local file name (to be written in download or to be read in upload). | Remote file name (to be written in upload or to be read in download). |
| The number of parallel channels for xFTSM and xPathM modes. | Protocol version and a unique session identifier specified in GUIDs. |
| TCP window size in bytes. | Desired block size of the used underlying storage system in bytes. |
| User credentials. | Extended mode (such as xDFS zero-copy parameters and etc.). |

TABLE 3: SOME TYPES OF CHANNEL EVENT

| Type | Description |
|---|---|
| **EOFT** | End of file reached and the session must be terminated by closing all channels. |
| **EOFR** | End of file reached in that channel but it must change its state to reusable channel mode. |
| **xFTSMU** | Initiate or change to xFTSM upload channel mode. |
| **xFTSMD** | Initiate or change to xFTSM download channel mode. |
| **xPathM** | Initiate or change to xPathM channel mode. |
| **NOOP** | No operation command over the channel. |
| **CONM** | Continue and maintain the latest channel event state. |
| **ZxDFS** | This channel is negotiating with remote channel in the zero-copy version of xDFS channels. |

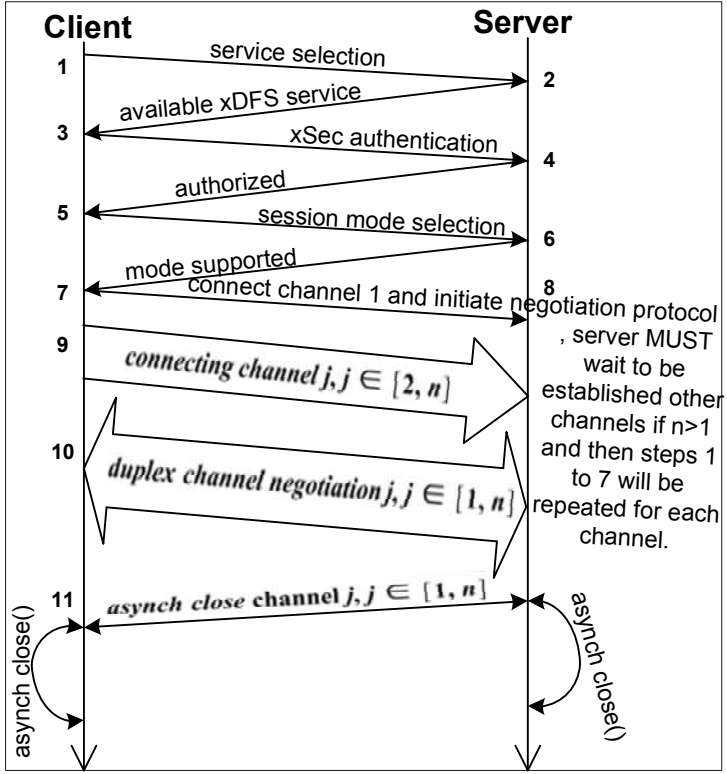
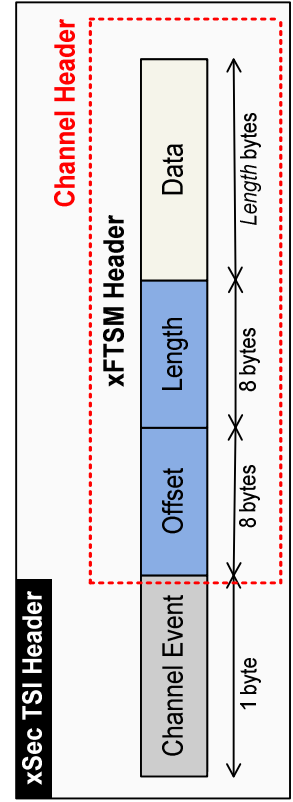

Fig. 4. Client-server xDFS protocol sequence diagram in xFTSM and xPathM modes, *n* is the number of parallel channels.

Fig. 5. A general xDFS protocol channel binary header encapsulated in xSec TSI header.

## 4. The Native, Cross-Platform and Cross-Language Implementation of xDFS Protocol

In this section a concrete explanation of the new xDFS implementation in native code is given atop xDotGrid.NET framework.

### 4.1. The Architecture of xDFS Implementation in Download and Upload Mode

The main core of the xDFS implementation is inspired from C# source codes of the DotDFS protocol. Anywhere it has been required according to the philosophy of the xDotGrid project, the C#-based DotDFS codes have been mapped into C++ codes. This sample mapping is important to prove the preliminary goal of the xDotGrid.NET framework for an application program in the real world. Fig. 6 portrays a simplified mapping of the C# xDfsClient class into the correspondent ISO C++ class. As seen in this figure due to the innate correspondency between the structural syntax of the C++ and C# languages, xDotGrid.NET makes a suitable framework for programmers to develop software systems in native C++ environment. As stated in Section 2.5.3, the MTEDP model suggests a general anatomy for xDFS server. In practice there does not exist an executable process file named the xDFS server notwithstanding. When xDotGrid service is executed on each of which network nodes are functioning, an executable file is run, and at any time needed in demand it instantiates necessary services and executes them. Fig. 7 shows the integrated hybrid xDotGrid server architecture. Each instance of xDotGrid server is comprised of at minimum six runtimes; e.g. in Fig. 7, there are three runtimes, including xThread Runtime, Common Runtime, and xFTSM Runtime. In this structure, *Listener Thread* (LT) receives the client requests and looks for what kind of service has been requested from the server through the transferred header over the *xDotGrid Socket Interface* (XDSI) channels. xThread is a new xDotGrid service, akin to the DotThreading model from DotGrid platform [6][10][11][12], which allows computational threads to be distributed across Grid nodes. Common Runtime has a number of duties such as monitoring of all running threads and the used physical memory by the clients requests, load-balancing between the execution cycles of threads in the entire CPU cores, and eventually the complete management of a full set of xDotGrid services running at server side.

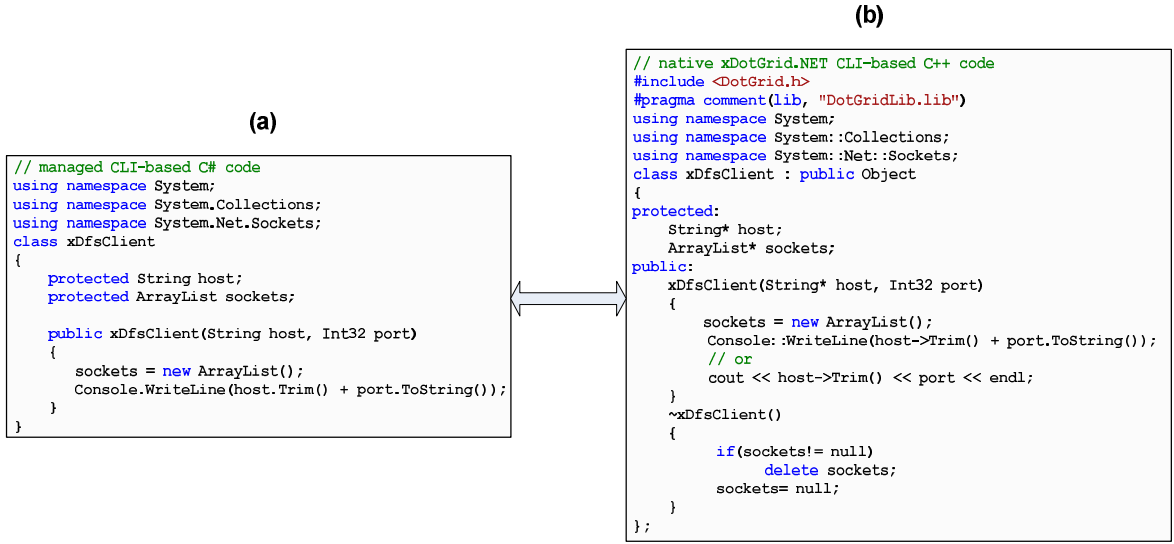

Fig. 6. Direct mapping between C#-based and native C++-based DotDFS/xDFS source codes. (a) Managed C# CLI code, (b) Native xDotGrid.NET C++ CLI-based code.

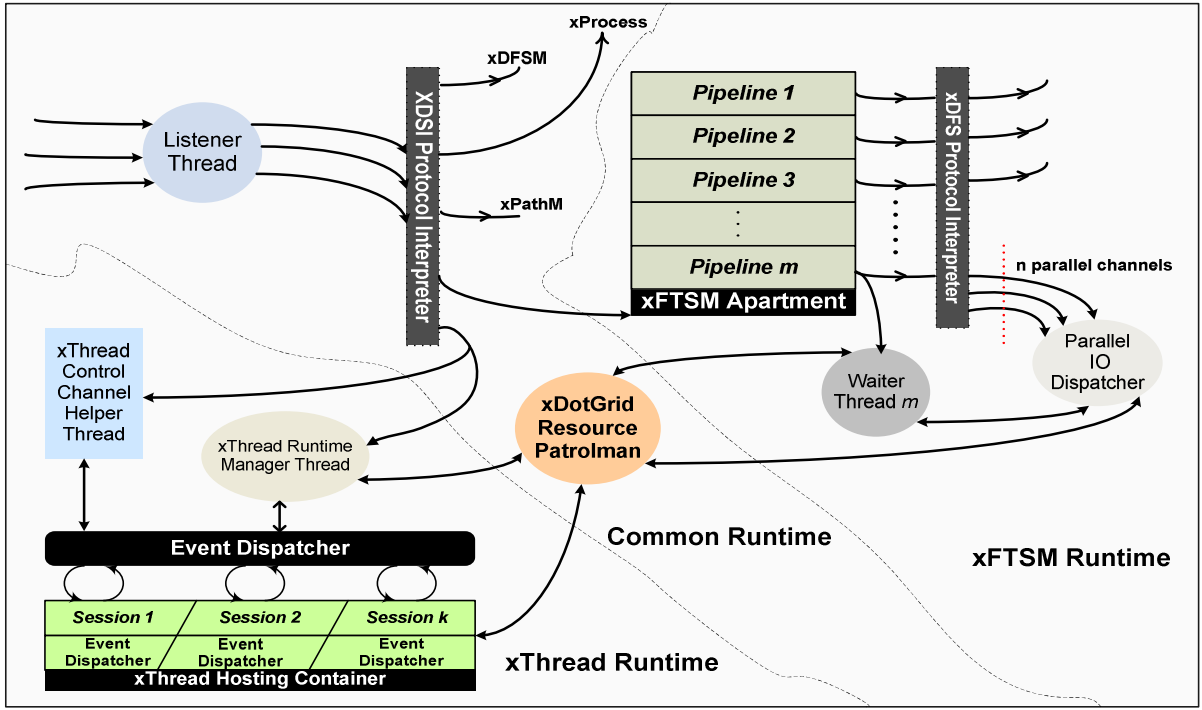

Fig. 7. Integrated hybrid xDotGrid server architecture.

TABLE 4: THE NUMBER OF THREADS

$$T_{hybrid} = 3 + m + \sum_{i=1}^{k}(S_i + 1) = 3 + m + \frac{k \cdot (k+1)}{2} + \sum_{i=1}^{k} S_i \quad (1)$$

xFTSM Runtime is the same implementation of the MTEDP model for file transfers in download and upload mode. Each pipeline in xFTSM Runtime is actually a container for one xFTSM session which is processing *n* parallel channels. xFTSM Apartment also manages *n* parallel channels. Waiter Thread is the thread which is run by LT to manage the execution flow of each service on demand. Each collection of the *n* parallel channels is simultaneously processed by a special module called as the *Parallel I/O Dispatcher* (PIOD). PIOD implements a C++ class interface in where one can extend its kernel relied on an extensive set of network event-dispatching mechanisms discussed in Section 2.5.3. PIOD, according to the type of upload or download mode, transfers the data packets between a client and server through asynchronous disk and network I/O methods. In Fig. 7, if we assume that Session $k$ has $S_k$ local threads then because Common Runtime and xThread Runtime, and xFTSM Runtime contain respectively three and $m$ threads, so the total number of threads in a xDotGrid server's instance can be calculated from the relation 1 of Table 4 at any moment of time. To fully understand the architectural implementation of xDotGrid server in xFTSM mode and more detail on that way which xDFS protocol functions in each either modes of client-server upload or download, we use CFSMs. Four corresponding CFSMs are illustrated in Figs. 8 to 11. For two reasons, one the download mode is usually used on the Internet and the other the page limitation of this paper, we only describe the client-server CFSMs of xDFS protocol in this section for download mode. In Fig. 8 after authenticating the client through xSec GSI, choosing the xFTSM mode by the client, and receiving the xFTSM parameters, the server checks whether the session has already been created using its GUID by the client or not.

If the session has already been existed and the number of sockets in the hash table is not corresponding (equal) to the value of *n* received from the client, the server adds the new client stream to the hash table in state 8. In step 7, the server concurrently checks, so that if the number of client streams is equal to the value of *n* then moves the CFSM flow to state 9. If an error occurs during states 1 to 8, the next state will be 18. Since the system function *select()* has been used as the event-dispatching component in the present xDFS design, these CFSMs were drawn depending upon the properties of this system routine. xDFS protocol has considered an *Exception Header* packet for the response of each request at any given side; in fact, this mechanism is a part of the concept of the duplex channel negotiation. If a client-side error takes place during any point of the file transfer session, then this header contains some binary details from an instance of the class *System::Exception* to be sent to the server through XDSI and *xDotGrid Remoting Architecture* (XDR), and the server decides how to deal with this error. In an erroneous point, the server either closes the current channel or terminates the entire transfer session. To preserve the state of sockets (for read-readiness and write-readiness modes), two individual array lists are used. Those sockets that are involved at the read-readiness list have three states labeled as *Done*, *NotDone* and *FirstTime*. The state *Done* interprets that the *Exception Header* value still hasn't been received by the server and the server in its next loop iteration, which manages the event-dispatching part, will have to receive this value from the client. The state *FirstTime* means that this socket for the first time has been used to be checked of its *Exception Hea*der. In Fig. 8, both socket lists are empty when the CFSM state changes from 7 to 9. In state 10, the event-dispatcher module is made up of two sub-parts each of which acts as nonblocking. The first and second event-dispatcher module belongs to the read-readiness and write-readiness list of sockets, respectively. The second event-dispatcher after filing the write-readiness socket list makes PIOD send the file blocks to the client. In the current xDFS implementation, the disk I/O management is performed in two ways as synchronous (blocking) and asynchronous (nonblocking). The traditional system procedures *read()* and *write()* are used for the synchronous mode, and an additional thread and one circular buffer are employed to implement the asynchronous core as well as these two routines.

To make use of a ring buffer and no use of system-level nonblocking mechanisms, like POSIX AIO, help maintain the xDFS portability across various platforms. The asynchronous disk I/O feature makes feasible the xDFS framework, without manipulating the xDFS implementation architecture, to be easily used in high throughput, low latency, quality of service and failover communications links, e.g. an InfiniBand interconnect, in where the disk speed bottleneck must be detached from the actual network throughput. PIOD drives out the write-readiness sockets from the write-readiness list after sending file blocks to them and then puts them into the read-readiness list in state 12. The state of the read-readiness list is changed to *NotDone* in state 12, so that the second event-dispatcher can check the *Exception Header* from the client side. If the end of file is reached, in step 15, the event-dispatcher checks whether the sent packets within the socket TCP buffer of the write-readiness list had been delivered to the client or not. If all file packets were sent to the client side, the server sends the end of file header to the whole connected client channels. If all headers were sent then the CFSM ends the current xFTSM session in transition from state 17 to 18. Fig. 9 portrays the client-side xDFS CFSM in xFTSM download mode. As it can be seen the CFSM in Fig. 9 is much simpler and this simplicity in client side is due to the order of data flow from server side to client in download mode. The states 1 through 5 are performed for all of the *n* parallel channels. Each client authenticates itself to the server and sends its related session information to server through *Download Request Header* over XDSI after channel establishment. This header contains the information shown in Table 2.

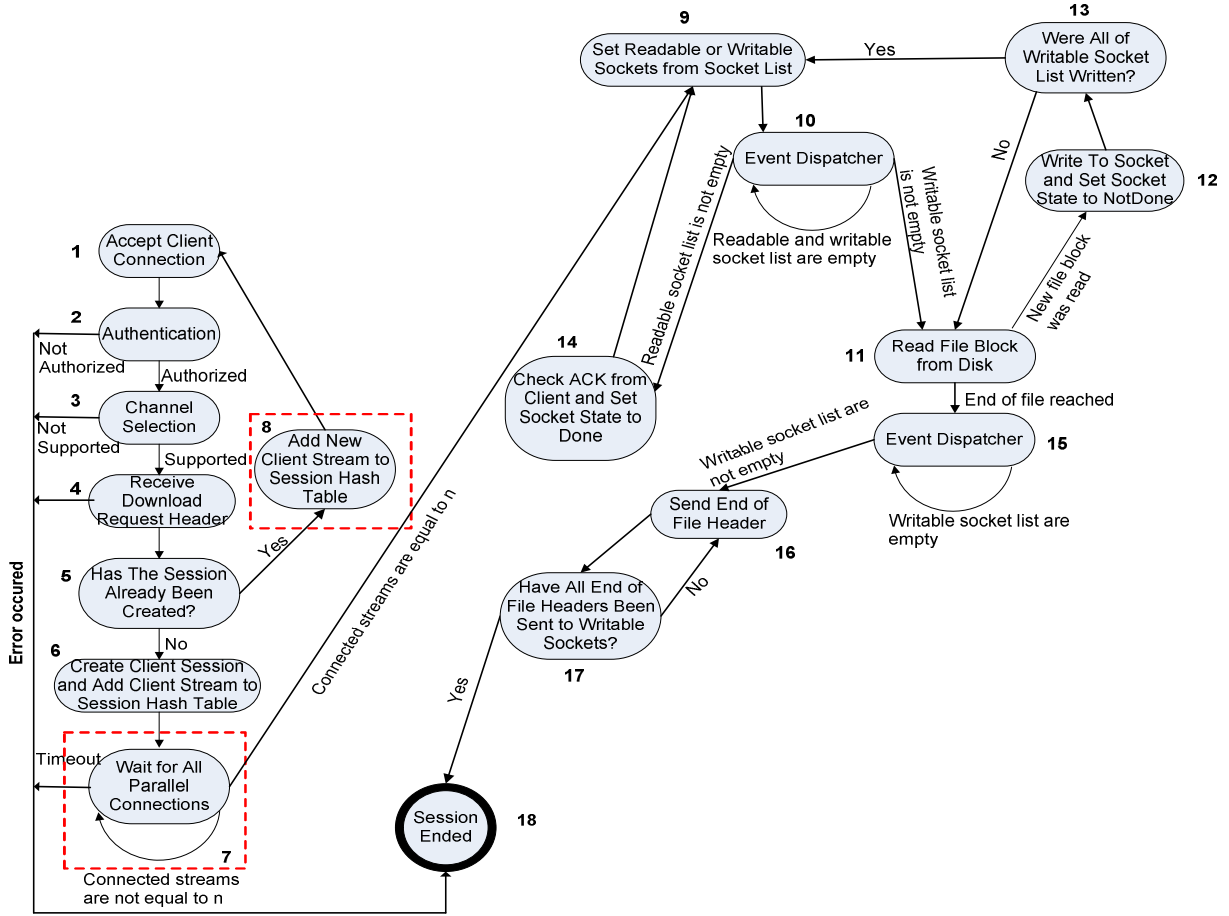

Fig. 8. xDFS server communicating finite state machine in xFTSM download mode.

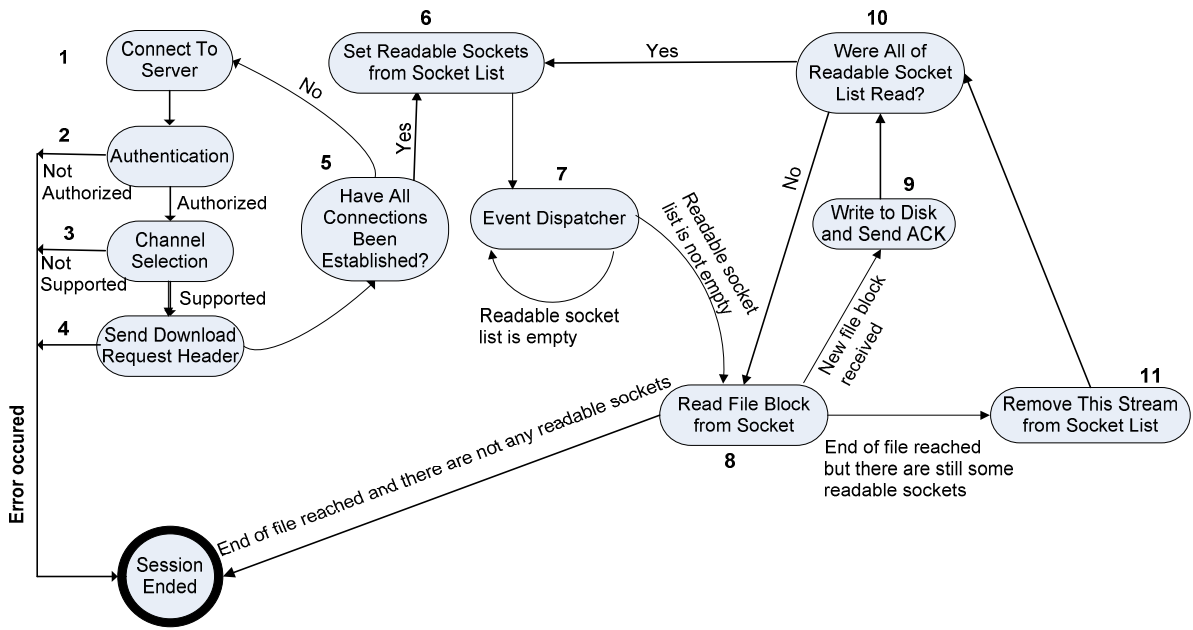

Fig. 9. xDFS client communicating finite state machine in xFTSM download mode.

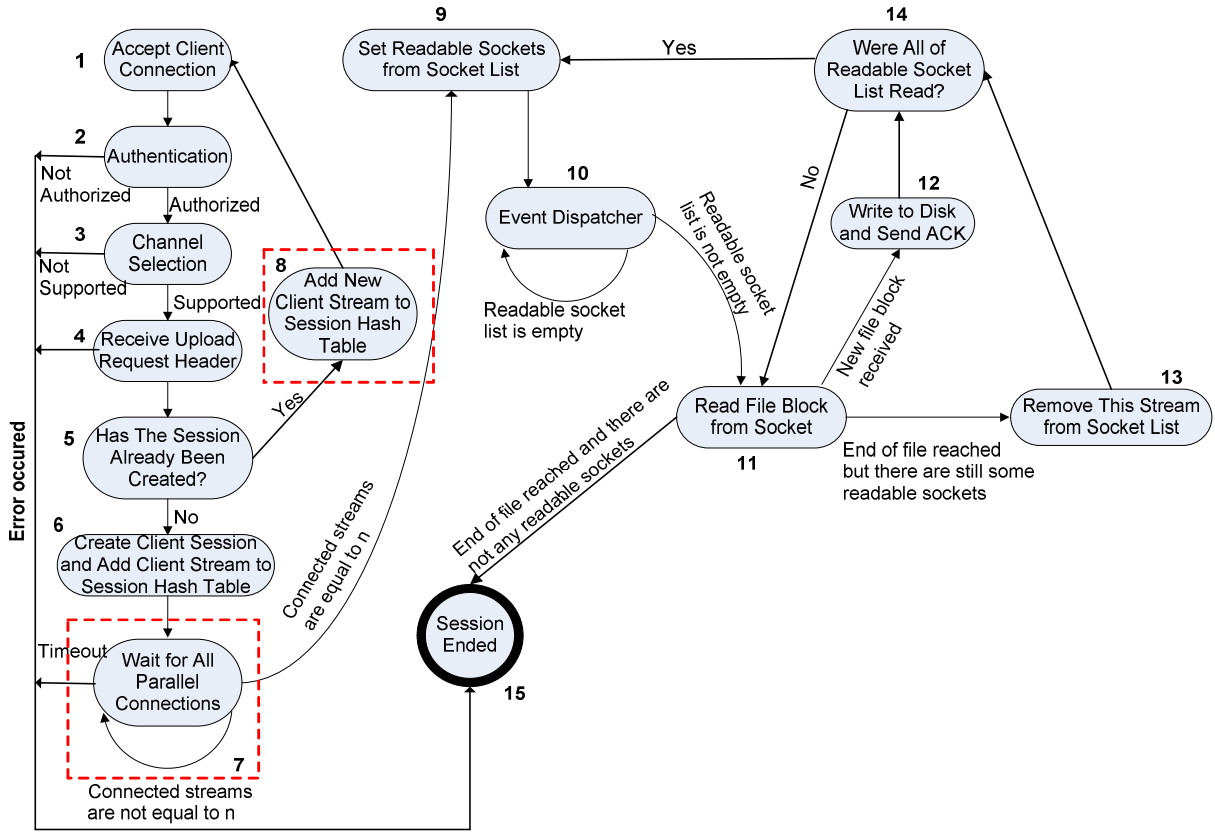

Fig. 10. xDFS server communicating finite state machine in xFTSM upload mode.

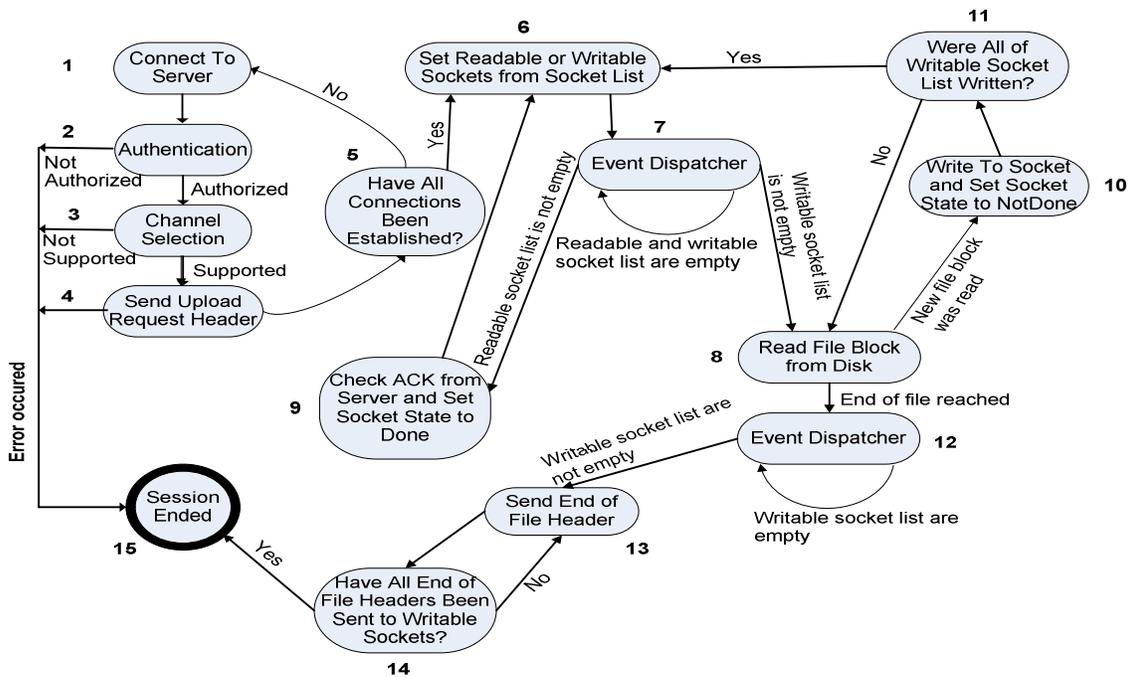

Fig.11. xDFS client communicating finite state machine in xFTSM upload mode.

The CFSM state changes from 5 to 6 after all channels were established. Inasmuch as TCP sockets are normally nonblocking at every sending operation, a write-readiness socket list has not been used in the CFSM of Fig. 9. In state 7, having chosen those sockets that infold the file blocks received from the server side, the event-dispatcher varies the CFSM state to 5. After being written the file blocks to the storage system through xDotGrid.NET Framework APIs, if all read-readiness sockets still haven't been written to the disk, the state 8 is repeated for them after the state 10. In step 8, if the server has already sent the end of file header to the client, then the CFSM steps into the state 12. In this state, the file transfer session in download mode terminates. By comparing the all CFSMs, it can be inferred that the right-hand side of server CFSMs in one mode has a one-to-one correspondence with the right-hand side of client CFSMs in another mode. Duality principle is referred to such a case in mathematics and graph theory. The deep review of this concept is out of the scope of this paper.

## 5. Experimental Studies

The current xDFS implementation only supports xFTSM mode with types of download and upload transfers. The real execution environment of the xDFS framework is native and cross-platform, and we tested and deployed the xDFS stably on a broad spread of operating systems, including UNIX, Linux, and Windows. To test and evaluate the real performance of the xDFS implementation in native code, xDFS is compared with the Globus GridFTP (explicitly the only available fully GridFTP implementation in UNIX-style operating systems) in a local-area network (LAN). The experimental results presented here have the confirmation on the logic of our previous works [7][8][9], but what's more they reveal interesting insights that are analyzed in this section. The test set is performed to characterize the xDFS throughput and efficiency in three categories: disk-to-disk, memory-to-memory and CPU/physical memory usage. The tests were done in a LAN network with 0.1 milliseconds (msec) round trip time (RTT) and a 1Gb/s bottleneck link. The machines used as the clients and servers had eight homogenous Intel Xeon Quad Core processors operating at 2.5GHz with 6MB cache, 8GB RAM and 320GB RAID hard disks. Linux CentOS with the kernel 2.6.9.9-78, for x86_64 SMP processors, was installed on all machines. The TCP buffer size and disk block size were set to 1MB. In GridFTP tests, its implementation of the GT4.2.1 was used [24]. The GridFTP C source codes were compiled into machine code by a makefile through the GNU GCC compiler suite. The xDFS C++ source codes have been constructed atop xDotGrid.NET framework. The xDotGrid.NET codes with an extension of one static library or dynamic shared library were implemented as cross-platform using the C-language preprocessors relied upon the core of Win32 and POSIX APIs. The xDotGrid project makes extensive customized use of the comprehensive cross-platform Code::Blocks IDE. Developers and programmers can easily and quickly implement, compile, debug and deploy their HPC/distributed applications based on the xDotGrid framework in a highly cross-platform/portable integrated environment. Finally, the C/C++ codes of the xDFS framework, through Code::Blocks and invoking the GCC compiler, are compiled and transformed into the native machine code. In the xDFS client side, the x-drotgrid-url-copy (XDUC) utility is used. As a unique feature of the xDotGrid platform, it should be noted that we manually ported the C#-based codes of the DotDFS's XDUC, which contained approximately 1000 lines of code, into the native ISO C++ code in less than 15 minutes (the XDUC itself instantiates and invokes the xDFS framework's APIs)! In the GridFTP client side, the GT4.2.1 globus-url-copy (GUC) was used [24]. All test points are a mean of 15 runs.

### 5.1. Single Stream Performance in Download Mode

We carried out the first comparison between xDFS and GridFTP in single stream test for download mode. Since there are two threads of execution for xDFS and four processes for GridFTP in client-server sides, this experiment can give a good benchmark to compare them just in the protocol level except for how these two protocols have been implemented (e.g., using a threaded-based or multiple-process implementation). Because both protocols have been implemented in native code, at the first look, it may hint that these protocols should have correspondent throughputs in single-stream scenarios; however, the shown results in Figs. 12 and 13 promote a quite distinct fact to the reader. Fig. 12 depicts throughputs for files of sizes ranging from 400MB to 4000MB. Furthermore, Fig. 13 demonstrates the percentage of client-server CPU usage for each protocol and the mentioned files. As it is obvious, the xDFS throughput at least 150Mb/s is better than GridFTP for files with sizes less than 1GB. In [7], we discovered a phenomenon that was defined as to the term *saturation speed*. There, we concluded that the saturation speed decreases the measured throughput to very low thresholds in contrast to the actual speed of the local reads and writes in storage systems under experiment when large files were being transferred. In Fig. 12, the saturation speed for GridFTP occurs while the file size is increased greater than or equal to 2 GB. Here clearly no saturation speed occurs for xDFS, as there was not observed any fall or decrease in xDFS throughput for large files. Whereas GridFTP forks just four processes in this case, we cannot consider the use of multiple processes as the key factor that reduces the throughput. A GridFTP server is consisted of three components, including GridFTP protocol module, the data transform module, and the Data Storage Interface (DSI) [25]. The GridFTP protocol module is the main module that performs the network send/receive operations and implements the protocol. This module has been built using the Globus eXtensible Input/output (XIO) [26][27]. The XIO is an OCRW (Open/Close/Read/Write) abstraction layer that simplifies the development phase of transport protocols. The XIO's architecture is comprised of two abstract concepts, drivers and stacks. In fact, the specifications of a protocol are included in a driver as abstract. Each driver must implement a set of well-defined function interfaces based upon the C-language *typedef* (as structure of a function pointer table) along with a collection of operations to enable dynamic runtime routine invocations. Extensively using function pointers to call functions, as XIO does, may produce a slow-down for the code on modern processors, because branch prediction may not be able to figure out where to branch to (it depends on the value of the function pointer at runtime) although this effect can be overstated as it is often amply compensated for by significantly reduced non indexed table lookups. There are two drivers, transform drivers and transport drivers. Transport drivers are those that actually convey data toward inside or outside the space of a process. As seen, XIO adds an extra abstraction layer between an application program and low-level system APIs (this abstraction is not created as static at compile time, rather is managed and processed dynamically at runtime). Then, these additional layers can cause overheads on architected programs in terms of systemic characteristics and network throughput.

Globus states just the main reason in using XIO for extensibility aspects and therefore GridFTP is able to leverage in order to be transport protocol agnostic. Hence, in environments where it makes sense, protocols much more aggressive than TCP can be utilized. To meet more specific extensibility needs, they also provide easy-to-use development libraries. They, in [28], states that achieved respectively 990Mb/s and 950Mb/s of the bottleneck bandwidth for Iperf and XIOPerf implemented based on XIO in a network with a bottleneck link of 1Gb/s. But Figs. 12 and specifically 13 show that GridFTP in contrast to xDFS exposes far more overheads due to using the XIO. It implies that the Globus shouldn't have used the XIO in designing very sensitive and important components such as GridFTP. In [28] they specify the overheads concerned with XIO increases linearly with incrementing the number of drivers. A GridFTP server or client typically comprised of three TCP, GSI and disk drivers. It is necessary noting that the overheads of a networked software system cannot be evaluated just depending on the network throughput, rather characterizing a variety of system characteristics have more importance, including cache misses and CPU utilization, interrupt handling, stalls, memory fetches, data locality, cache utilization and thread interactions, and so on. Moreover, the GridFTP uses the DSI [25] in access to the functionality of storage systems such as file systems accessible via standard POSIX API, and Storage Resource Broker. DSI abstraction provides a modular pluggable interface to the data storage systems. DSIs can be loaded and switched dynamically at runtime. When a GridFTP server needs the storage system, it passes a request to the loaded DSI instance. The DSI after servicing the request notifies the server from the service completion. In contrast to the Globus XIO and Globus DSI, the xDFS framework makes use of the *xDotGrid Socket Interface* (XDSI) and *CLI FileStream Interface* (CLIFSI) when the integrated access to the network I/O interfaces and heterogeneous storage systems. XDSI and CLIFSI are actually two pure C++ classes that inherit from the interface (abstract/base) class *System::IO::Stream* and implement it. Virtual methods are not used to implement sensitive method stubs like *Read()* and *Write()* of this base class. A virtual call requires at least an extra indexed dereference, and sometimes a fixup addition, in contrast to a non-virtual call, which is simply a jump to a compiled-in pointer. Hence, calling virtual functions is inherently slower than calling non-virtual functions. Experiments verify that approximately 6-13% of execution time is spent simply dispatching to the correct function where the overhead can be as high as 50% [29]. These two classes have been carefully designed relied on the xDotGrid.NET inline expansion methods. Therefore, the XDSI and CLIFSI as static binding are distributed across the codes of the xDFS framework through inlining mechanisms by the compiler. Since these procedures are static, and we here avoid the use of any virtual methods, the XDSI and CLIFSI expose no overhead on the xDFS implementation. The XIO internally uses a number of event synchronization on the stacks to ensure the users, using the library, that they are receiving events in a reasonable state. For example, a barrier is used between all the data operation events. So the rationale of the saturation speed phenomena can be induced to the Globus DSI and XIO, and with the mechanisms of event notifications employed in implementing both of them. Globus considered just overheads originated from XIO over the throughput [26][27][28]; however, Fig. 13 reveals other technical facts. In Fig. 11 for this test, XIO poses at least 20 percent of the overhead in CPU usage in compared with xDFS particularly for larger files; also this percentage linearly increases for GridFTP, as it is invariable for xDFS and even degrades in some cases.

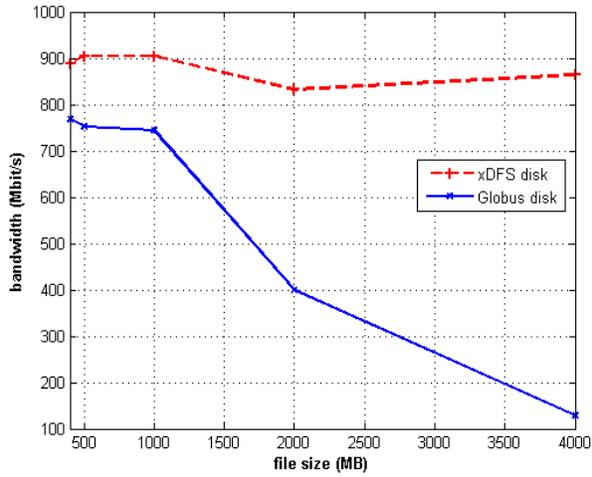

Fig. 12. Single stream throughput in download mode.

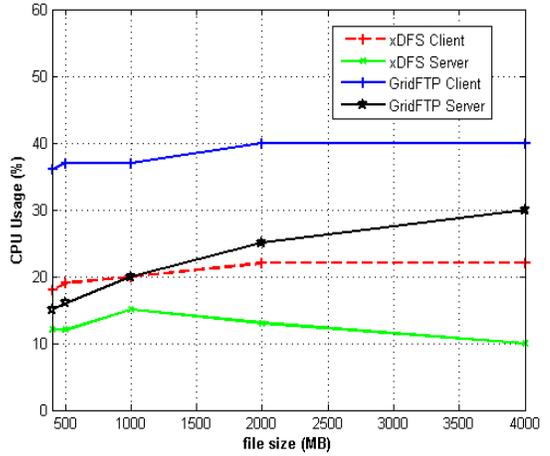

Fig. 13. Client-server CPU usage for different file sizes in single stream download mode.

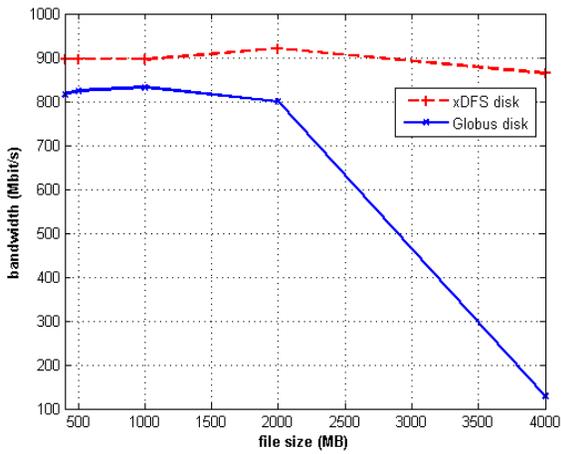

Fig. 14. Single stream throughput in upload mode.

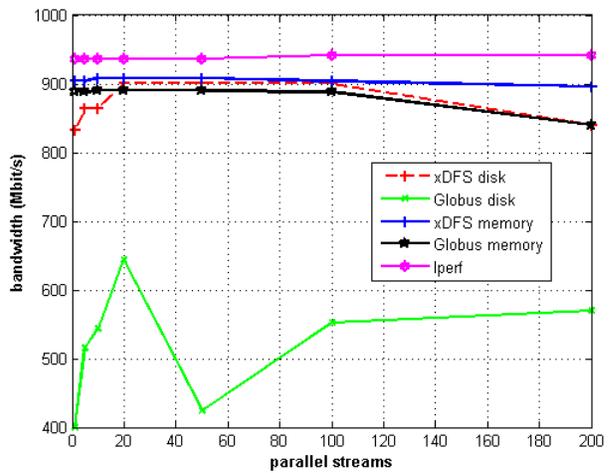

Fig. 15. Parallel throughput in download mode.

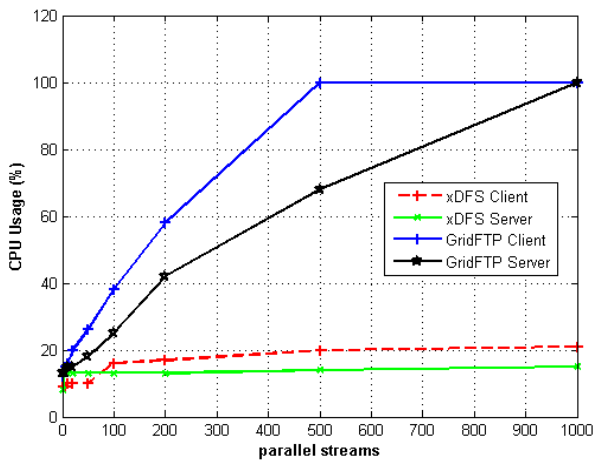

Fig. 16. Client-server CPU usage for memory-to-memory tests in download mode.

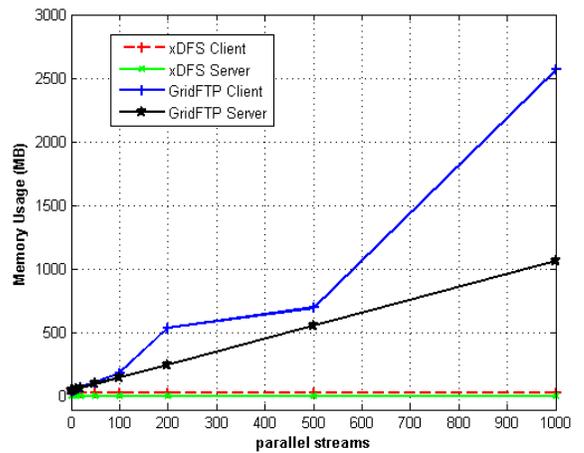

Fig. 17. Client-server memory usage for disk-to-disk transfer of a 4GB file in download mode.

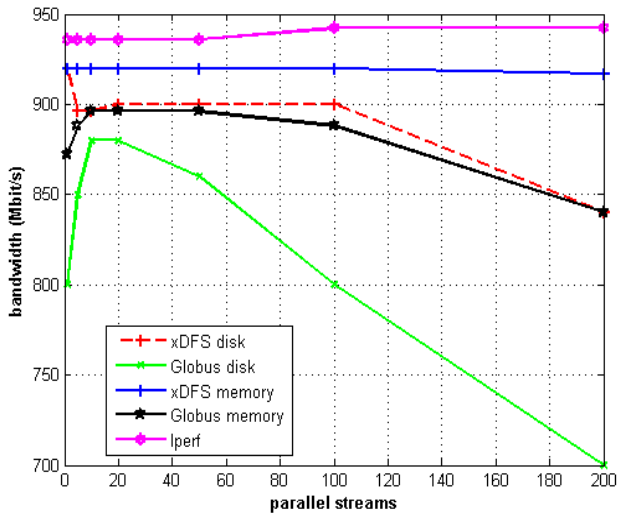

Fig. 18. Parallel throughput in upload mode.

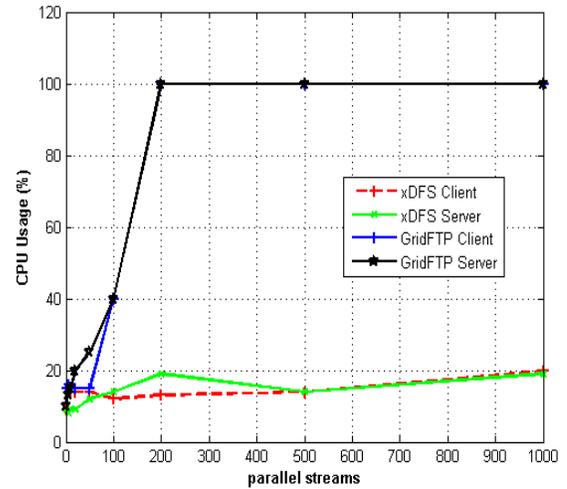

Fig. 19. Client-server CPU usage for memory-to-memory tests in upload mode.

### 5.2. Single Stream Performance in Upload Mode

The second experiment is performed for single stream in upload mode for the two protocols xDFS and GridFTP. Fig. 14 shows the throughputs. The xDFS upload-mode profile is virtually identical to its download-mode profile, but these profiles for GridFTP differ. Because both Figs. 12 and 14 relate to the single-stream throughputs, the main factor for being xDFS throughputs alike and being GridFTP throughputs dissimilar can be attributed to the different structure and implementation of the Globus XIO and DSI for GET/PUT modes in GridFTP.

### 5.3. Harnessing Parallelism in Download Mode

In this section, we probe the effect of multiple streams on the overall throughput in download mode. This test is actually a touchstone to compare the MP model in Section 2.5.1 for GridFTP protocol and the MTEDP model in Section 2.5.3 for xDFS protocol. Fig. 15 depicts the obtained throughputs as a function of the number of parallel streams in download mode. This data set was carried out for three different experiments: Iperf, memory-to-memory tests (/dev/zero to /dev/null) and disk-to-disk tests. A 2GB file is transferred between client and server in disk-to-disk tests. In memory-to-memory tests for download mode, the xDFS and GridFTP reached to the 97% and 95% of the bottleneck bandwidth, respectively. As it can be seen, because of using a single thread and event-driven methods, the memory-to-memory and disk-to-disk xDFS throughput is almost constant as for GridFTP it has very high fluctuations. Clearly, all of the xDFS throughputs are better than GridFTP in all cases. Section 2 discussed the implementation architecture of these two protocols. As stated earlier by increasing the number of processes in GridFTP structure the overhead associated with the protocol significantly increases, the fact that exhibits itself in Fig. 15. In the disk-to-disk tests here, the xDFS throughput at least 256 Mb/s and at most 432 Mb/s was superior to the GridFTP. An interesting point was to observe that the xDFS disk-to-disk profiles followed the xDFS memory-to-memory profiles with a very little difference in all tests. To reveal more overheads of the MP model upon GridFTP implementation for a wide range of the number of parallel streams, we measured the CPU usage for a memory-to-memory test and the physical memory usage for transferring a 4GB file. Figs. 16 and 17 show the result of these two experiments. While the number of parallel streams is increasing, the client-server CPU usage for the xDFS is constant and between 5 to 20%, as it is exponentially increasing for the GridFTP programs. In this scenario, both the percentage of CPU usage and the physical memory consumption for the GridFTP client are greater than the GridFTP server. The percentage of CPU usage of both the GridFTP server and client are 100% in 1000 parallel streams over the whole 32 CPU cores! As another conspicuous issue during our experiments in download mode, the Linux machine who hosted the GridFTP client became critically as unresponsive which made us manually reboot that machine for transferring large files (greater than or equal to 4 GB) with parallel streams greater than 200. Fig. 17 illustrates the physical memory usage in megabytes for xDFS and GridFTP programs. Obviously, the profile curve of xDFS physical memory consumption is a flat line in very little values; by contrast, this profile for GridFTP increasingly consumes the physical memory while raising the number of parallel streams. Based on technical points studied in this and previous section, the GridFTP due to its intrinsic architecture and using an improper implementation (a process-based implementation, the use of UNIX-based forks, and using the Globus DSI and XIO) suffers from critical overheads. Hence, GridFTP cannot be used in high-traffic and vital environments like Internet services, and data-intensive Grid and Cloud applications.

**5.4. Harnessing Parallelism in Upload Mode**

In the last experiment, we examine the effectiveness of parallel streams in upload mode for a 2GB file transfer based on the three tests conducted in Section 5.3. Figs. 18 and 19, for a set of wide range of parallel streams, respectively show the parallel throughput and the percentage of CPU usage for memory-to-memory tests. xDFS and GridFTP in these experiments reached to the 98.5% and 95% of the bottleneck bandwidth, respectively. In [7] we stated that DotDFS in such an experiment reached to the 94% of the bottleneck bandwidth and its implementation relied on .NET framework imposed at least 5% overhead on the throughput. In this experiment we witnessed that the native code performance with a peak of 98.5% of the bottleneck bandwidth was achieved by the xDFS framework. Another interesting and unique highlight is that as it can be seen from Fig. 18, the profiles of xDFS throughput in disk-to-disk and memory-to-memory overlap in the range of 5-10 of parallel streams with the peak bandwidth of 920 Mb/s. This fact engages two important points regarding the used event-driven model within the xDFS protocol. First, the use of an event-driven anatomy is crucial in xDFS and DotDFS protocols as two optimum file transfer protocols. Second as stated in [7], the DotDFS protocol is the first concurrent file transfer protocol that, with the integration of thread-based and event-driven models, proposes a new computing paradigm in the field of file transmission protocols. In upload mode with parallel streams, it seems that the overall GridFTP throughput improves beside the download mode. The origin of this variation might be attributed to the mechanisms used in the design of Globus XIO and Globus DSI. In various disk-to-disk tests for GridFTP in upload mode we found out a profile of parabolic curves as a concave function of the number of parallel streams. One example of these parabolic curves is shown in Fig. 18. The profile has a global maximum point at the 20 value of parallel streams as increasing these streams lead to the dramatic decrease in throughput. The CPU-usage profiles in Figs. 16 and 19 indicate that the percentage of CPU utilization (consumption) in the server side of a protocol for upload mode has a duality form in relation to its client side for download mode and vice versa. The percentage of CPU consumption among all CPU cores reaches to the 100-percent utilization while the number of parallel streams is increasing greater than (or equal to) 200.

**6. The Position of the xDFS File Transfer Framework among Parallel Storage Systems**

Over the last decade, especially in recent years, a great importance has been given to the data transfer problem, particularly for data-intensive applications, because data is the only way in the realization of the whole computing. Different parallel storage systems have been designed and implemented to cover specific requirements; in this regard, we first discuss some of them and then explain the common and novel position of our xDFS framework among the broad spectrum of such these systems. *Lustre* is a shared parallel file system that is usually used for large-scale cluster computing [30]. In this system, a single metadata server (MDS) stores the namespace metadata like file names, directories, access information, and file layouts. File data are stored and retrieved on one or more object storage servers (OSSes). Clients can simultaneously have the read and write access permissions through the standard POSIX semantics. Lustre layer network (LNET) supports different network interconnects such as InfiniBand and TCP/IP over the Ethernet. Lustre is available only for Linux platforms. *HDFS* (Hadoop Distributed File System) is a parallel distributed file system, which got inspired from Google File System, written in the non-native Java language for Hadoop framework [31]. In HDFS, every data node distributes the data blocks across the network relied upon a block protocol specific to the HDFS through the TCP/IP layer and RPC to communicate with client agents. HDFS is not a fully POSIX-compliant file system. Access to files in HDFS can be provided through native Java API. *GFS* (Google File System) is a proprietary distributed file system that was designed for use in the Google's web search platform [32]. It provides reliable and efficient access to data over a large cluster of commodity hardware. GFS splits huge files into 64-megabyte chunks, and the most files are mutated only with appending the data to their endings. *GPFS* (General Parallel File System) is a proprietary shared-disk cluster file system that is owned by IBM [33]. GPFS achieves higher I/O performance through stripping the data blocks on multiple disks, and the read and write operation of these blocks in parallel. GPFS fully supports the POSIX semantics and is designed for high-end hardware not for the commodity hardware. In contrast to parallel file systems above, cloud storage systems have been getting a key role of the Cloud technology in recent few years. Different providers and cloud infrastructures are releasing new storage services from time to time. Despite the performance maturity reached within the area of parallel distributed file systems, it has not yet been achieved completely in cloud services and thus there is enough academic space to research about cloud storage infrastructures. Amazon Simple Storage Service (Amazon S3) [34] and Microsoft Windows Azure Cloud Storage [35] are from two major cloud storage systems. Amazon S3 is an object storage provided on the cloud that has a simple web services interface to store and retrieve data wherever through the Web. Objects are fundamental entities stored in Amazon S3 which can contain up to 5 TB of data. Microsoft is one of the biggest cloud storage providers. The most basic cloud storage of this provider is Azure Blob that provides object storage on the cloud. Azure Blob is a distributed storage system for large data items. Each item can be up to 50 GB in size. In reviewing all the above systems, we can arrive at some common points that determine the position of the xDFS framework as an infrastructural service. We survey the most important of them. Communication is one of the important issues in the design of a parallel storage system. Most systems make use of RPC to communicate that in turn it enables them to execute independent of the underlying operating systems, networks and protocols. Thus, the network independence is necessary to support heterogeneous networks. Security, especially cloud environments in where one user accesses to storage systems through the Web medium, is the second key issue. Most systems achieve in their own security by leveraging authentication, authorization and privacy of the existing security systems, which this could cause critical problems for specific applications. The cross-platform issue and synchronization semantics in accessing to data are two other features that must be considered by parallel storage systems. Most parallel file systems are designed only for a single platform or operating system that limits their use in other platforms. In addition, providing the synchronization POSIX semantics in concurrent access of clients to shared data through file locking mechanisms is very important.

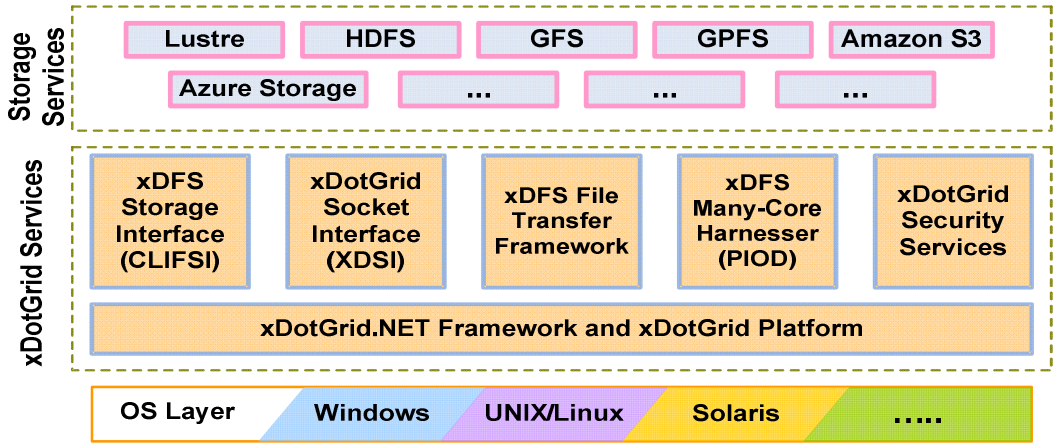

Fig. 20. The key xDotGrid services so as to develop parallel and cloud storage systems.

Now, we discuss the position of the xDFS framework among parallel storage systems. With the outlined material, Fig. 20 is derived. Several systems as mentioned earlier locate at the top layer *Storage Services*. The middle layer *xDotGrid Services* serves five services to its upper layer for managing and transferring data, which it instead allows the top layer (parallel storage systems) to operate on a wide variety of heterogeneous platforms and operating systems. All these services were discussed in this paper. The CLIFSI service helps its upper layer systems benefit from the functionality of various hardware storage systems (e.g., traditional hard disks or SAN storage) for manipulating physical data, files and directories. The XDSI service ensures network-independence property and allows the upper systems to use any type of the underlying communication protocols for data exchange (such as WAN, InfiniBand, Ethernet, and so forth). Particularly that XDSI is a good alternative for RPC and RMI, because it supports data serialization/deserialization and provides certain methods for cross-process invocation/intercommunication across any network of computers. The xDFS file transfer framework along with three xFTSM/xDFSM/xPathM modes, which establish a high throughput and performance protocol, are fundamental to all systems within the layer *Storage Services* to manage and transfer data over all networks, including LAN, MAN, and WAN. The PIOD service, which integrates both thread-based and event-driven models into one hybrid pattern, takes advantage of executing many concurrent transactions in order to harness the power of massively task parallelism on many-core processors, which can be used by parallel storage systems its above. Since the xDFS framework has been developed based on xSec security layer, which itself has been designed with the Grid security demands in mind, it can seamlessly satisfy the data transfer security and integrity, and the authentication and authorization of remote storage entities. All of these services have been implemented atop the native xDotGrid.NET framework that allows the exploitation of native code performance, platform independence, and portability heavily. Although the xDFS framework can be arranged in stripped data transmission scenarios similar to the existing parallel file systems [6][7][11][10][11], but its main strength does not appear for current and future storage systems at first glance. It lets storage system programmers avoid unnecessary details of the low-level systems programming and rather focus on their own system. Consequently, the code development cycle will be enhanced extremely, and developers can utilize the broad benefits of xDotGrid platform.

**7. Conclusion and Future Works**

In this paper, we described the new xDFS file transfer protocol and its new extensions on DotDFS protocol. Also, we examined the development methods of optimal file transfer systems, and their advantages and disadvantages. Architectural differences between two protocols xDFS and GridFTP were considered, by which we proposed the xDFS protocol as a basis for high performance/throughput file transfers in data-intensive Grid/Cloud/Internet environments. With introducing the xDFS implementation atop xDotGrid.NET framework [6], it was illustrated to the reader how we can make use of the CLI-set standards relied upon the xDotGrid platform to develop a standardized distributed software infrastructure as a general example. The presented results confirmed the accuracy of appropriate methods used in the design and implementation of the xDFS protocol. The xDFS framework is now available as a cross-platform and native-code distribution for a wide family of operating systems like UNIX, Linux and Windows. In all disk-to-disk tests for transferring a 2GB file with or without parallelism, the xDFS throughput at least 30% and at most 53% was superior to the GridFTP. Memory-to-memory tests in upload mode showed that the xDFS protocol accessed to the 98.5 percent of the bottleneck bandwidth while the GridFTP protocol was reaching to the 95%. We have determined the preliminary roadmap of our future research works based on what was presented in this article. Currently, we are finalizing the specification drafts of the xDFSM and xPathM protocols and subsequently to implement them within the xDFS framework. Due to the heavy design period of the xDotGrid project, another work will be to port the DotSec protocol into native code. After completion of the xDFS framework we intend to extend the PIOD architecture discussed in Section 4 so that several network I/O event-dispatching approaches can be used for the sake of more data transfer efficiency.

**Acknowledgments**

The author, A. Poshtkohi, would like to thank gratefully from his parents (Mr. Abdullah Poshtkohi and Ms. Effat Soltanieh) that extremely helped him in his life create this work while he had been growing in the past ten years after starting his academic studies.

**Alireza Poshtkohi** received the B.Sc. and M.Sc. degrees in Electrical and Electronic Engineering respectively from Islamic Azad University of Qazvin, Qazvin, Iran in 2006, and Shahed University, Tehran, Iran in 2011. He has authored an English textbook in computer science and several journal articles. His current research interests include operating systems, distributed/parallel/Grid/Cloud computing, network protocols, highly concurrent systems, programming languages, in-kernel software abstractions, sandboxing virtual machines, and Grid-based low-power circuit design methodologies.

**M.B. Ghaznavi-Ghoushchi** (M'07) received the B.Sc. degree from the Shiraz University, Shiraz, Iran, in 1993, the M.Sc. and Ph.D. degrees both from the Tarbiat Modares University, Tehran, Iran, in 1997, and 2003 respectively. He is currently an Assistant Professor with Shahed University, Tehran, Iran. His current research interests include VLSI Design, Low-Power and Energy-Efficient circuit and systems, Computer Aided Design Automation for Mixed-Signal and UML-based designs for SOC and Mixed-Signal.